\documentclass[
prb,
twocolumn,
amsmath,amssymb,aps,nobibnotes,
superscriptaddress,
notitlepage,
nofootinbib,
preprintnumbers
,floatfix
]{revtex4-2}

\usepackage{graphicx}
\usepackage[colorlinks=true,linkcolor=blue,urlcolor=blue,citecolor=blue]{hyperref}
\usepackage{subcaption}
\usepackage{bm}
\usepackage{braket}
\usepackage{mathtools}
\usepackage{siunitx}
\usepackage{enumitem}

\graphicspath{{figures/}}

\newcommand{\kondo}{J_{\mathrm{K}}}
\newcommand{\heisen}{J_{\mathrm{H}}}

\begin{document}

\title{Cavity control of quantum phase transitions in a two-dimensional Kondo lattice}

\author{Jun Mochida}
\email{jun-mochida@g.ecc.u-tokyo.ac.jp}
\affiliation{Department of Physics, University of Tokyo, 7-3-1 Hongo, Bunkyo-ku, Tokyo 113-0033, Japan}
\author{Ata\c c $\dot{\mathrm{I}}$mamo$\breve{\mathrm{g}}$lu}
\affiliation{Institute of Quantum Electronics, ETH Zurich, CH-8093 Z{\"u}rich, Switzerland}
\author{Yuto Ashida}
\affiliation{Department of Physics, University of Tokyo, 7-3-1 Hongo, Bunkyo-ku, Tokyo 113-0033, Japan}
\affiliation{Institute for Physics of Intelligence, University of Tokyo, 7-3-1 Hongo, Tokyo 113-0033, Japan}

\date{\today}

\begin{abstract}
Cavity quantum electrodynamics offers a route to control quantum phases by using vacuum fluctuations of confined electromagnetic fields.
In particular, planar cavities based on polar van der Waals materials can generate strongly confined modes and are promising for controlling two-dimensional correlated materials.
Recently, moir\'e materials have become central platforms for studying two-dimensional heavy-fermion systems and their quantum phase transitions.
Kondo lattices provide a prototypical model for studying quantum phase boundaries, driven by competition between Kondo screening and the ordering of local magnetic moments.
We show that a cavity-induced interaction can shift the quantum phase transitions between a heavy-fermion phase and an antiferromagnetic phase in a two-dimensional Kondo lattice through a momentum-dependent self-energy of the conduction bands.
For the longitudinal projected field motivated by h-BN hyperbolic phonon polaritons, the self-energy favors Kondo hybridization and expands the heavy-fermion region.
Transverse and circular in-plane model structures give distinct effects, with the transverse case relatively favoring the magnetically ordered phase and the circular case lying between the longitudinal and transverse cases.
These results indicate that electromagnetic vacuum fluctuations can effectively modify the control parameters of strongly correlated two-dimensional Kondo materials.
\end{abstract}

\maketitle

\section{Introduction\label{sec:intro}}
Controlling quantum materials with strong vacuum fluctuations of electromagnetic fields is one of the promising directions in condensed matter physics.
In recent years, advances in cavity engineering have paved the way for manipulating electronic states of matter through ultrastrong coupling between electrons and a quantized electromagnetic environment~\cite{forn-diazUltrastrongCouplingRegimes2019,schlawinCavityQuantumMaterials2022,blochStronglyCorrelatedElectron2022,luCavityEngineeringSolidstate2025}.
Ultrastrong coupling achieves the regime where the light-matter coupling strength $g$ becomes comparable to the bare electronic energy scales such as THz frequencies.
Vacuum fluctuations in the ultrastrong regime, virtual photon processes, and photon-mediated interactions can modify electronic properties even in the ground state.
This means that the cavity vacuum field can be used to control equilibrium properties, without applying a classical external driving to the material.
Recent experiments have indeed reported cavity modifications of superconductivity~\cite{thomasExploringSuperconductivityStrong2025a,kerenCavityalteredSuperconductivity2026}, metal-insulator transitions~\cite{jarcCavitymediatedThermalControl2023}, and breakdown of topological protection in quantum Hall systems~\cite{appuglieseBreakdownTopologicalProtection2022}.

A particularly interesting direction is the application of this idea to strongly correlated systems.
In parallel, numerous experimental and theoretical studies have explored ferroelectricity~\cite{ashidaQuantumElectrodynamicControl2020}, topological phenomena~\cite{rokajLightMatterInteraction2018,dmytrukGaugeFixingStrongly2021,masukiBerryPhaseTopology2023,dagEngineeringTopologyGraphene2024,guerciSuperradiantPhaseTransition2020,gomez-leonMajoranaBoundStates2025,enknerTunableVacuumfieldControl2025}, and a variety of strongly correlated phenomena, including superconductivity~\cite{liManipulatingIntertwinedOrders2020,gaoHiggsModeStabilization2021,talkingtonUltrastrongCouplingSignatures2026}, magnetism~\cite{roman-rochePhotonCondensationEnhanced2021,sentefQuantumClassicalCrossover2020,chiocchettaCavityinducedQuantumSpin2021,fadlerEngineeringPhotonmediatedLongrange2024,weberCavityrenormalizedQuantumCriticality2023,masukiCavityMoireMaterials2024}, non-Fermi-liquid states~\cite{raoNonFermiLiquidBehaviorCavity2023,surAmplifiedResponseCavitycoupled2025,guoCavityInducedExcitonicInsulation}, the Kondo effect~\cite{kuoKondoQEDKondo2023,mochidaCavityenhancedKondoEffect2024}, and related phenomena~\cite{garcia-vidalManipulatingMatterStrong2021,paravicini-baglianiMagnetotransportControlledLandau2019,kippCavityElectrodynamicsVan2025a,bartoloVacuumdressedCavityMagnetotransport2018,debernardisCavityQuantumElectrodynamics2018a,friskkockumUltrastrongCouplingLight2019,leboiteTheoreticalMethodsUltrastrong2020,ashidaCavityQuantumElectrodynamics2021,ashidaNonperturbativeWaveguideQuantum2022,andolinaTheoryPhotonCondensation2020,rokajPolaritonicHofstadterButterfly2022,saez-blazquezCanWeObserve2023,enknerTestingRenormalizationKlitzing2024,nakamotoOnedimensionalExtendedHubbard2025,kassManyBodyPhotonBlockade2024,debernardisLightmatterInteractionsVacuum2024,nambiarDiagnosingElectronicPhases2025,kimSymmetryControlledUltrastrongPhononPhoton2025,graziottoCavityQEDControl2025,helmrichCavitydrivenAttractiveInteractions2026}.

To date, discussions of the ultrastrong coupling regime have mainly focused on collective electronic excitations in which the light-matter coupling strength $g\propto \sqrt{N}$ scales with the number $N$ of emitters coupled to a mode of the cavity, similar to the Dicke model.
For two-dimensional materials, planar-cavity structures based on polar van der Waals (vdW) layers such as h-BN support strongly confined modes whose properties depend on the layer thickness~\cite{ashidaCavityQuantumElectrodynamics2023,herzigsheinfuxHighqualityNanocavitiesMultimodal2024a,guoHyperbolicPhononpolaritonElectroluminescence2025}.
In the THz band, hyperbolic phonon--polaritons can confine light at deeply sub-wavelength scales, suggesting the feasibility of realizing ultrastrong coupling for a single electron.
Such cavity platforms might make it possible to explore novel quantum phases in two-dimensional materials under the ultrastrong coupling regime, which is not accessible in bulk materials.

In heavy-fermion materials, the interplay between localized magnetic moments and a sea of itinerant electrons leads to rich many-body phenomena.
These materials, typically represented by a Kondo lattice model, play an important role in many topics of condensed matter physics, including the relation to unconventional superconductivity and quantum criticality~\cite{hewsonKondoProblemHeavy1993,colemanKondostabilisedSpinLiquids1989,colemanIntroductionManyBodyPhysics2015}.
The Kondo lattice model shows the competition between the heavy-fermion phase emerging from the Kondo effect and the magnetic order.
The antiferromagnetic (AFM) exchange interaction between the local moments and conduction electrons leads to spin screening, called the Kondo effect.
When this local singlet formation becomes coherent over the lattice, massive quasiparticles called heavy fermions emerge.
On the other hand, conduction electrons also contribute to the magnetic ordering between local moments driven by the Ruderman-Kittel-Kasuya-Yosida (RKKY) interaction~\cite{doniachKondoLatticeWeak1977,tsunetsuguGroundstatePhaseDiagram1997,hewsonKondoProblemHeavy1993,petersFerromagneticStateOnedimensional2012,colemanIntroductionManyBodyPhysics2015,minamiLowTemperatureMagneticProperties2015,zhouDoniachPhaseDiagram2023}.
An especially intriguing class of quantum critical points (QCPs) in heavy-fermion systems involves the breakdown of the Kondo effect, where the screening of local moments is destroyed, leading to the abrupt reconstruction of the Fermi surface~\cite{siLocallyCriticalQuantum2001,colemanHowFermiLiquids2001,senthilWeakMagnetismNonFermi2004,colemanTransportAnomaliesSimplified2005,shenStrangemetalBehaviourPure2020,paschenQuantumPhasesDriven2021,sachdevQuantumPhasesMatter2023}.
The Kondo-screened phase has a large Fermi surface, while a Kondo-broken local-moment phase has a small Fermi surface.

This type of QCP and transition has been suggested to occur in several materials.
For instance, recent transition-metal dichalcogenide moir\'e MoTe$_2$/WSe$_2$ experiments have observed signatures of Kondo breakdown, showing that gate-tunable moir\'e Kondo lattices can access quantum phase transitions related to Kondo breakdown, strange metal behavior, and topological phases~\cite{kumarGatetunableHeavyFermion2022,guerciChiralKondoLattice2023,zhaoGatetunableHeavyFermions2023,zhaoEmergenceFerromagnetismOnset2024,xieKondoEffectIts2024}.
More broadly, in the area of two-dimensional materials, moir\'e materials exhibit a wide range of rich physical properties~\cite{bistritzerMoireBandsTwisted2011,andreiMarvelsMoireMaterials2021,dalalOrbitallySelectiveMott2021,ramiresEmulatingHeavyFermions2021,songMagicAngleTwistedBilayer2022}, and among them, they have opened a new route to heavy-fermion systems and their QCP physics.
In particular, moir\'e Kondo lattice systems are attractive because the Kondo coupling and carrier density can be tuned by gates, and because the triangular moir\'e geometry naturally introduces magnetic frustration~\cite{dalalOrbitallySelectiveMott2021,kumarGatetunableHeavyFermion2022,guerciChiralKondoLattice2023,zhaoGatetunableHeavyFermions2023,zhaoEmergenceFerromagnetismOnset2024}.
Experiments have demonstrated gate-tunable heavy fermions in MoTe$_2$/WSe$_2$ moir\'e bilayers~\cite{zhaoGatetunableHeavyFermions2023,zhaoEmergenceFerromagnetismOnset2024,hanTopologicalKondoInsulator2026,hanStackingorderdependentElectronicProperties2026}.
This combination makes the moir\'e platform suitable for studying the competition between magnetic order and the Kondo effect.
The continuous Kondo-breakdown QCP scenarios motivate this competition, but the transition obtained below from our restricted mean-field comparison is first order and is not a direct description of a continuous Kondo-breakdown QCP.

\begin{figure}[t]
  \centering
  \includegraphics[width=\columnwidth]{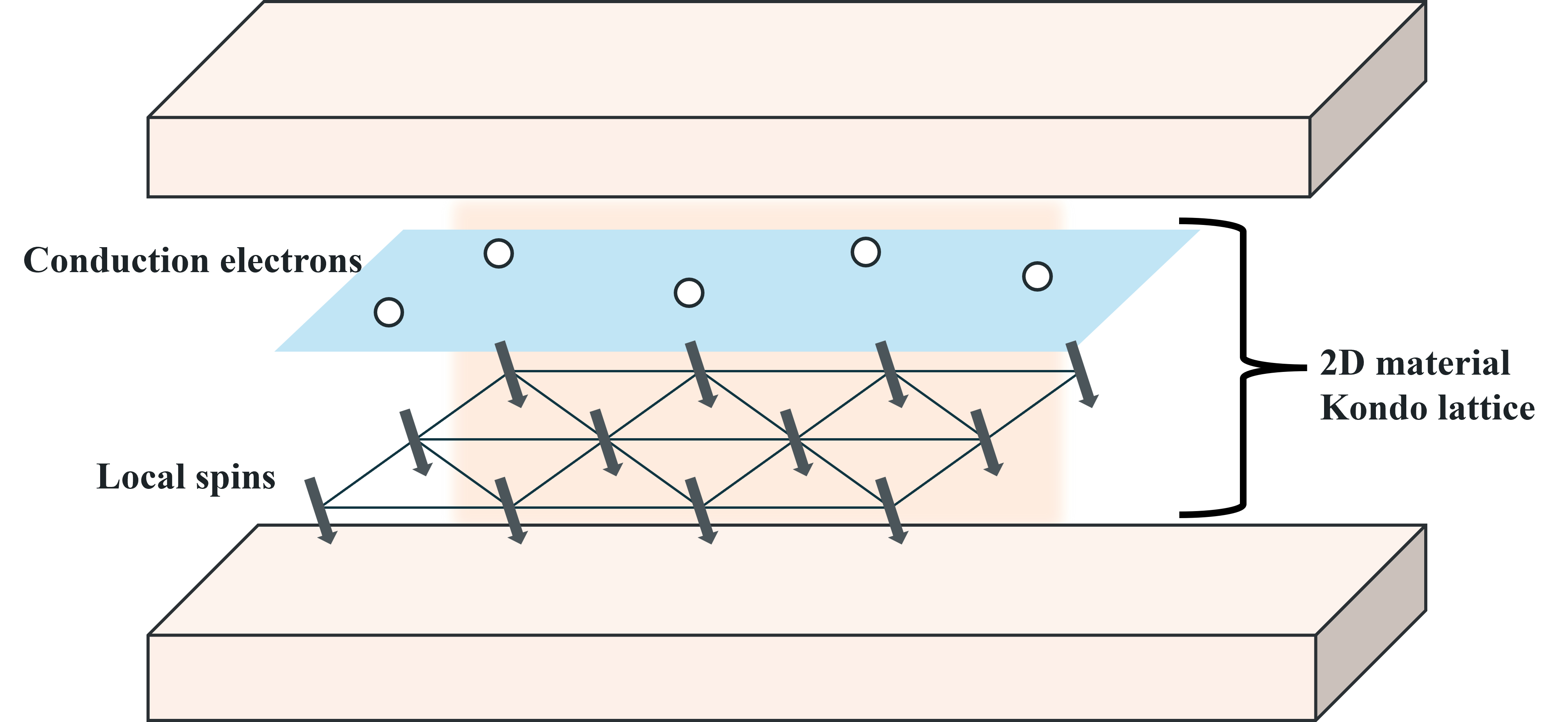}
  \caption{Schematic setup of a two-dimensional Kondo lattice material coupled to a planar cavity mode. Conduction electrons in the two-dimensional Kondo lattice layer couple to the cavity mode, while the localized moments are treated as decoupled from the cavity field.}
  \label{fig:cavity_setup}
\end{figure}

In this paper, we study how cavity vacuum fluctuations can modify the competition between Kondo screening and magnetic order in a two-dimensional Kondo lattice  (Fig.~\ref{fig:cavity_setup}).
Starting from a periodic Anderson model coupled to cavity modes in two-dimensional materials, we derive the leading cavity-induced correction to the conduction-electron band and include it in a mean-field treatment of a triangular Kondo lattice.
We focus on the triangular Kondo lattice as a minimal model motivated by moir\'e heavy-fermion platforms, where carrier density, Kondo coupling, and magnetic frustration can be tuned.
This analysis allows us to identify how cavity-induced modifications of the conduction-electron band change the competition between the heavy-fermion and magnetic phases.

We note that the cavity effect on a single magnetic impurity has been studied in Ref.~\cite{mochidaCavityenhancedKondoEffect2024}.
There, the effective interaction induced by the cavity gives an effect similar to mass enhancement of conduction electrons.
As a result, the density of states near the Fermi surface is effectively enhanced, and the Kondo temperature increases.
However, it remained unclear how this mechanism works in a lattice model, where the same conduction electrons also mediate magnetic correlations between local moments, and the cavity can affect not only Kondo screening but also magnetic order.

The rest of the paper is organized as follows.
In Sec.~\ref{sec:model}, we introduce the cavity Kondo lattice model and the mean-field methods used in this study.
In Sec.~\ref{sec:results}, we present the cavity-induced self-energy, the phase diagram for the AFM and the heavy-fermion phases at a metallic filling, and a low-density comparison between the ferromagnetic (FM) phase and heavy-fermion phases.
In Sec.~\ref{sec:discussion}, we discuss how the self-energy affects the phase transition and give a physical interpretation of its polarization dependence.
In Sec.~\ref{sec:conclusion}, we summarize the results of this paper.

\section{Model and Methods\label{sec:model}}
\subsection{Periodic Anderson Model Coupled to Cavity Modes\label{sec:model_eff}}
First, we consider a periodic Anderson model as a starting point for a heavy-fermion system.
The conduction electrons are coupled to the cavity modes through the minimal coupling.
In the present model, we assume that Zeeman-type coupling to the localized moments is much smaller than the coupling to conduction electrons in the THz regime, so the localized $f$ electrons are assumed to be decoupled from the cavity modes.
In this work, we consider using cavity modes supported by a planar cavity consisting of thin polar van der Waals crystals~\cite{ashidaCavityQuantumElectrodynamics2023}.
The h-BN hyperbolic phonon-polariton modes provide the concrete motivation for the longitudinal projected in-plane field discussed below.

We start from the Coulomb-gauge minimal-coupling Hamiltonian of the periodic Anderson model,
\begin{align}
  H&= H_{\mathrm{el}} + H_{\mathrm{\text{el--ph}}} + H_{\mathrm{cav}}, \notag\\
    H_{\mathrm{el}} &=
    \sum_{\bm{k},\sigma}\frac{\hbar^2\bm{k}^2}{2m}c^\dagger_{\bm{k}\sigma}c_{\bm{k}\sigma}
    +\sum_{i,\sigma} \varepsilon_f f^\dagger_{i\sigma}f_{i\sigma}
    \notag\\
    &+U\sum_i n^f_{i\uparrow}n^f_{i\downarrow}
    +\sum_{i,\sigma}\left(Vc^\dagger_{i\sigma}f_{i\sigma}+\mathrm{H.c.}\right), \notag\\
    H_{\mathrm{\text{el--ph}}} &= \frac{e}{m}\sum_{\sigma}\int d^2\bm{r}\,
    c^\dagger_{\sigma}(\bm{r})
          \bm{A}(\bm{r})\cdot\bm{p}\,c_{\sigma}(\bm{r}) \notag\\
    &\quad+\frac{e^2}{2m}\sum_{\sigma}\int d^2\bm{r}\,
    c^\dagger_{\sigma}(\bm{r})c_{\sigma}(\bm{r})\bm{A}^2(\bm{r}),
    \notag\\
    H_{\mathrm{cav}} &=\sum_{\bm{q}}\hbar\omega_{\bm{q}}a^\dagger_{\bm{q}}a_{\bm{q}},
    \label{eq:pa_model}
\end{align}
where $H_{\mathrm{el}}$ is the periodic Anderson Hamiltonian without the cavity field, $H_{\mathrm{\text{el--ph}}}$ is the light-matter coupling term, and $H_{\mathrm{cav}}$ is the free cavity Hamiltonian.
We note that $e$ is the electron charge, $m$ is the electron mass, $\hbar$ is the reduced Planck constant, $\bm{p}$ is the momentum operator for the conduction electrons and $\bm{r}$ is the coordinate operator.
$c_{\bm{k}\sigma}$ and $f_{i\sigma}$ are the annihilation operators for conduction electrons with momentum $\bm{k}$ and spin $\sigma$ and localized electrons at site $i$ with spin $\sigma$, respectively for spin-$1/2$ fermions, $a_{\bm{q}}$ is the annihilation operator for the cavity mode with momentum $\bm{q}$.
$\varepsilon_f$ is the energy level of the localized $f$ electrons, $U$ is the on-site Coulomb repulsion for $f$ electrons, $V$ is the hybridization strength between conduction and localized electrons, $n^f_{i\sigma}=f^\dagger_{i\sigma}f_{i\sigma}$ is the number operator of the localized $f$ electron, and $\hbar\omega_{\bm{q}}$ is the energy of the cavity photon mode.

The in-plane vector potential is given by
\begin{align}
    \bm{A}(\bm{r}) =
    \sum_{\bm{q}}A_{\bm{q}}\left[
      \bm{e}_{\bm{q}}a_{\bm{q}}e^{i\bm{q}\cdot\bm{r}}
      +\bm{e}_{\bm{q}}^\ast a^\dagger_{\bm{q}}e^{-i\bm{q}\cdot\bm{r}}
    \right].
    \label{eq:vector_potential}
\end{align}
The in-plane vector $\bm{e}_{\bm{q}}$ denotes the component of the cavity mode projected onto the two-dimensional material plane.
The real coefficient $A_{\bm{q}}$ denotes the amplitude of the vector potential for momentum $\bm{q}$.
For an effective field confinement $L^2$, we use the usual plane-wave mode normalization in which $A_{\bm{q}}^2\propto L^{-2}$.
In an h-BN hyperbolic cavity, for example, the field coupled to the two-dimensional electrons is longitudinal after projection onto the material plane \cite{ashidaCavityQuantumElectrodynamics2023}.
The electromagnetic modes satisfy the three-dimensional transversality condition in the Coulomb gauge.
After projecting the physical cavity modes onto the two-dimensional electron layer, their in-plane components can be parallel to the in-plane momenta.
Thus, the modes in the h-BN setup are expected to have effectively momentum-parallel in-plane projections, which we call longitudinal.
Correspondingly, one can also consider a perpendicular in-plane structure, which we call transverse.
The same physical modes can also be formulated in the generalized Coulomb gauge, $\nabla\cdot[\epsilon(\bm{r})\bm{A}(\bm{r})]=0$, where $\epsilon(\bm{r})$ is the spatially dependent dielectric function.
In the vacuum region relevant for the electron layer, this condition reduces to the usual transversality condition, while the dielectric environment determines the profile and normalization of each mode.
We regard the longitudinal projected field as the setup motivated by h-BN hyperbolic phonon polaritons, while the transverse and circular in-plane structures considered below are model extensions used to diagnose how the momentum structure of the cavity-induced self-energy affects the Kondo-lattice phase competition.

We treat the cavity interaction by perturbation theory with the interaction parameter $\eta$, following the single-band formulation of Ref.~\cite{masukiCavityMoireMaterials2024}, which is controlled at least in the $\eta\ll1$ regime.
To this end, we assume the cavity state to be in the vacuum state and project the full Hamiltonian onto the photon vacuum.
This projection justifies the perturbative treatment of the light-matter coupling.
Following the perturbation theory summarized in Appendix~\ref{app:effective_interaction}, we obtain the effective interaction between electrons mediated by the cavity photons in the form
\begin{align}
    H_{\mathrm{ee}} &=
    \sum_{\bm{q}}\frac{\hbar^2g_{\bm{q}}^2}{m\omega_{\bm{q}}^2}
    \sum_{\bm{k},\bm{k}',\sigma,\sigma'}(\bm{e}_{\bm{q}}\cdot\bm{k})(\bm{e}_{\bm{q}}^\ast\cdot\bm{k}')
    \notag\\
    &\quad\times
    c^\dagger_{\bm{k}+\bm{q},\sigma}c^\dagger_{\bm{k}'\sigma'} c_{\bm{k}'+\bm{q},\sigma'}c_{\bm{k}\sigma}.
    \label{eq:electron_electron}
\end{align}
This cavity-induced electron--electron interaction implies that the cavity mediates a long-range interaction between electrons whose momentum-dependent weight depends on the projected in-plane polarization structure.
Here, $g_{\bm{q}}\equiv eA_{\bm{q}}\sqrt{\omega_{\bm{q}}/(m\hbar)}$ is the effective light-matter coupling strength of the cavity mode with momentum $\bm{q}$.
A dimensionless parameter $\eta$ is used to characterize the overall strength of the cavity coupling and is defined as $\eta^2=\sum_{\bm{q}}g_{\bm{q}}^2/\omega_{\bm{q}}^2$,
where the sum runs over the cavity modes retained in the effective model.
The interaction in Eq.~\eqref{eq:electron_electron} is kept as the leading cavity effect on the conduction electrons.

For large $U$ in the local-moment regime, the periodic Anderson model reduces to a Kondo lattice model with an exchange interaction $\sim V^2/U$ at leading order in the hybridization.
By contrast, the direct cavity correction to $\kondo$ appears only in higher order in both the light-matter coupling and the hybridization.
Therefore, we neglect this direct correction and include the cavity effect mainly through the conduction-electron self-energy generated by Eq.~\eqref{eq:electron_electron}.

\subsection{Cavity-Induced Self-Energy and Effective Kondo Lattice\label{sec:self_method}}
We next treat the cavity-induced interaction in Eq.~\eqref{eq:electron_electron} at leading Hartree--Fock order.
For the zero-current mean-field states considered here, the Hartree contraction is proportional to the conduction-electron current and vanishes.
We thus focus on the Fock contribution, as discussed in Appendix~\ref{app:effective_interaction}.
This gives a static and momentum-dependent correction to the conduction-electron band.

The self-energy for a wave vector $\bm{k}$ from the cavity interaction is written as
\begin{align}
    \Sigma(\bm{k}) =
    -2L^2\int\frac{d^2q}{(2\pi)^2} \frac{\hbar^2g_{\bm{q}}^2}{m\omega_{\bm{q}}^2}|\bm{e}_{\bm{q}}\cdot\bm{k}|^2 f(\varepsilon_{\bm{k}-\bm{q}}),
    \label{eq:self_energy_intro}
\end{align}
which is real, static, and non-positive in this approximation.
Here, $L^2$ is the area of the two-dimensional system, and $f(\varepsilon)$ is the Fermi distribution function.
The renormalized energy band of conduction electrons is given by $\varepsilon_{\bm{k}}=\varepsilon^0_{\bm{k}}+\Sigma(\bm{k})-\mu$, where $\varepsilon^0_{\bm{k}}$ is the bare dispersion without the cavity interaction and $\mu$ is the chemical potential.
In a realistic moir\'e material, ordinary Coulomb interactions and microscopic band-structure effects can already renormalize this low-energy dispersion, and such renormalization may depend on carrier density.
We assume that this ordinary Coulomb-induced band renormalization is phenomenologically incorporated into the reference dispersion $\varepsilon^0_{\bm{k}}$.
To isolate the cavity-induced modification of the conduction-electron band, we incorporate the Hartree--Fock self-energy $\Sigma(\bm{k})$ generated by the cavity-mediated interaction on top of this reference dispersion.
This procedure yields an effective periodic Anderson model with a cavity-modified conduction-electron dispersion.
The cavity Kondo lattice model is then derived by applying the standard Schrieffer--Wolff transformation in the local-moment regime with $V/U\ll1$, which gives
\begin{align}
    H_{\mathrm{cKL}} &=
    \sum_{\bm{k},\sigma}\varepsilon_{\bm{k}} c^\dagger_{\bm{k}\sigma}c_{\bm{k}\sigma}
    +\kondo\sum_i \bm{S}_i\cdot\bm{s}_c(\bm{r}_i),
    \label{eq:ckl}
\end{align}
where $\bm{s}_c(\bm{r}_i)=\sum_{\alpha\beta}c_{i\alpha}^\dagger(\bm{\sigma}_{\alpha\beta}/2)c_{i\beta}$ is the spin density of the conduction electrons at $\bm{r}_i$, $\bm{\sigma}$ denotes the Pauli matrices, and both the conduction electrons and the localized moments $\bm{S}_i$ are taken to be spin-$1/2$ degrees of freedom.
At the particle-hole symmetric point, $\varepsilon_f=-U/2$, the Schrieffer--Wolff transformation gives the antiferromagnetic Kondo exchange $\kondo=8V^2/U>0$.
For sufficiently large $U/t$, this relation allows $\kondo$ to be of a few times $t$ while retaining $V/U\ll1$.
The cavity dependence enters Eq.~\eqref{eq:ckl} through $\varepsilon_{\bm{k}}$, while $\kondo$ is treated as an independent control parameter.
We denote the conduction-electron filling by $\rho_{\mathrm e}=N_{\mathrm e}/N_0$, where $N_{\mathrm e}$ is the number of conduction electrons and $N_0$ is the number of lattice unit cells.

In the Kondo lattice model, the conduction electrons mediate an RKKY interaction~\cite{aristovIndirectRKKYInteraction1997}.
In the present mean-field calculation we introduce a nearest-neighbor Heisenberg interaction,
\begin{align}
    H_{\mathrm{H}} = \heisen\sum_{\langle i,j\rangle}\bm{S}_i\cdot\bm{S}_j,
    \label{eq:effective_heisenberg}
\end{align}
as an independent phenomenological interaction.
Although the RKKY interaction is microscopically determined by the conduction-electron spin susceptibility, the present mean-field calculation does not determine this interaction self-consistently.
We therefore fix the Heisenberg interaction as an effective representation of the magnetic tendency~\cite{siLocalFluctuationsQuantum2003,senthilWeakMagnetismNonFermi2004,pixleyQuantumPhasesShastrySutherland2014,bernhardCoexistenceMagneticOrder2015,liPhaseEvolutionTwodimensional2015,wangQuantumPhaseTransition2020,sachdevQuantumPhasesMatter2023}.
In this sense, keeping $\heisen$ fixed while varying $\kondo$ and $\eta$ isolates how the cavity-induced conduction-band self-energy changes the relative stability of the magnetic and heavy-fermion phases.
For the triangular-lattice calculation at $\rho_{\mathrm e}=0.7$, we choose $\heisen>0$ and assume the 120-degree antiferromagnetic ordering.
At low electron density, we instead assume a fixed ferromagnetic interaction $\heisen<0$, motivated by the ferromagnetic tendency reported in Ref.~\cite{akagiHiddenMultipleSpinInteractions2012}.
A fully self-consistent treatment in which the cavity-modified susceptibility determines the magnetic interaction is left for future work.

\subsection{Mean-Field Treatment of Competing Phases\label{sec:mean_method}}
In order to compare the heavy-fermion phase and the magnetically ordered state, we use an Abrikosov fermion mean-field theory for the Kondo lattice model~\cite{newnsMeanfieldTheoryIntermediate1987,colemanKondostabilisedSpinLiquids1989,senthilWeakMagnetismNonFermi2004,
pixleyQuantumPhasesShastrySutherland2014,bernhardCoexistenceMagneticOrder2015,liPhaseEvolutionTwodimensional2015,colemanIntroductionManyBodyPhysics2015,sachdevQuantumPhasesMatter2023,kumarGatetunableHeavyFermion2022,guerciChiralKondoLattice2023}.
This method provides a common mean-field description of Kondo screening and local-moment magnetism.
First, we decompose the local spin moments into Abrikosov fermions,
$\bm{S}_i=f_{i\alpha}^{\dagger}(\bm{\sigma}_{\alpha\beta}/2)f_{i\beta}$.
Here, $f_{i\alpha}$ denotes an auxiliary fermion to represent the local spin and is not the physical localized-orbital electron $f_{i\sigma}$ of the periodic Anderson model in Eq.~\eqref{eq:pa_model}.
The Abrikosov fermion representation contains unphysical empty and doubly occupied states, and one has to impose the local constraint $\sum_{\alpha}f_{i\alpha}^{\dagger}f_{i\alpha}=1$ at each site.
This constraint is imposed by a Lagrange multiplier $\lambda$.
The interaction terms are then decoupled into mean-field parameters.
In the self-consistent mean-field calculations described below, we fix the conduction-electron filling $\rho_{\mathrm e}$ and determine $\mu$ self-consistently alongside the Lagrange multiplier $\lambda$.

The hybridization mean-field solution is the saddle point of the large-$N$ expansion and becomes exact for $N\to\infty$.
The present spin-$1/2$ calculation at $N=2$, with a restricted set of magnetic states, is a static mean-field approximation.
The coherent hybridized-band picture of the zero-temperature heavy-fermion ground state is qualitatively supported by studies beyond static mean-field theory, while the disappearance of the mean-field hybridization at finite temperature is a saddle-point artifact~\cite{beachCoherenceMetamagnetism2008}.
Our zero-temperature calculations focus on the cavity-induced shift of the transition point within this mean-field approximation.

\subsubsection{Heavy-Fermion Phase}

In the heavy-fermion phase, each localized magnetic moment tends to be screened by the surrounding conduction electrons through the Kondo effect.
When this screening becomes coherent over the lattice, the nearly flat $f$-fermion level hybridizes with the conduction band and forms narrow heavy quasiparticle bands.
As a result, the effective mass of the quasiparticles becomes much larger than the bare electron mass, and the low-energy quasiparticles are built from both conduction electrons and localized moments~\cite{colemanIntroductionManyBodyPhysics2015}.

The Kondo hybridization order parameter $P$ comes from the mean-field decoupling of the Kondo exchange, and it measures the mixing between conduction electrons and Abrikosov fermions.
Thus a nonzero $P$ describes the heavy-fermion phase.
The full mean-field Hamiltonian for the heavy-fermion phase is
\begin{align}
    H_{\mathrm{MF}}^{\mathrm{Kondo}} &=
    \sum_{\bm{k},\sigma}\varepsilon_{\bm{k}}
    c^\dagger_{\bm{k}\sigma}c_{\bm{k}\sigma} + \lambda\sum_{\bm{k},\sigma}
    f^\dagger_{\bm{k}\sigma}f_{\bm{k}\sigma} \notag\\
    &\quad
    +P\sum_{\bm{k},\sigma}
    \left(
      c^\dagger_{\bm{k}\sigma}f_{\bm{k}\sigma} + f^\dagger_{\bm{k}\sigma}c_{\bm{k}\sigma}
    \right)
    + N_0\left(\frac{2P^2}{\kondo}-\lambda\right).
    \label{eq:hf_mf_hamiltonian}
\end{align}
Here the last term is the mean-field constant term, including the local constraint for one $f$-fermion per site.
With this Abrikosov fermion mean-field theory, the mean-field Hamiltonian for each momentum is written as
\begin{align}
    H_{\bm{k}}^{\mathrm{Kondo}} =
    \begin{pmatrix}
        \varepsilon_{\bm{k}} & P \\
        P & \lambda
    \end{pmatrix},
    \qquad
    \varepsilon_{\bm{k}}=\varepsilon^0_{\bm{k}}+\Sigma(\bm{k})-\mu.
    \label{eq:Hk_kondo}
\end{align}

The Hamiltonian in Eq.~\eqref{eq:Hk_kondo} acts on the spinor $\Psi_{\bm{k}\sigma}^{\mathrm{Kondo}}=(c_{\bm{k}\sigma},f_{\bm{k}\sigma})^T$.
Here $\varepsilon_{\bm{k}}$ contains the cavity-induced self-energy and the chemical potential.
The Hamiltonian is spin degenerate, so the spin label is omitted.
A finite $P$ therefore incorporates the localized moments into the hybridized low-energy quasiparticle bands.
The two quasiparticle bands are denoted by $E_{\bm{k}}^{\pm}$, whose eigenvalues are obtained as
\begin{align}
    E_{\bm{k}}^{\pm} =
    \frac{1}{2}\left[
      \varepsilon_{\bm{k}} + \lambda
      \pm \sqrt{(\varepsilon_{\bm{k}}-\lambda)^2+4P^2}
    \right].
    \label{eq:kondo_eigenvalues}
\end{align}
The fixed-density mean-field energy of the heavy-fermion phase is
\begin{align}
    E_{\mathrm{MF}}^{\mathrm{Kondo}}
    &=\frac{2}{N_0}\sum_{\bm{k},\nu=\pm}
    E_{\bm{k}}^{\nu}f(E_{\bm{k}}^{\nu})
    +\frac{2P^2}{\kondo}-\lambda+\mu\rho_{\mathrm e}.
    \label{eq:energy_hf}
\end{align}
The self-consistent equations for the mean-field parameters
are obtained from the stationarity conditions of the corresponding mean-field functional.
At zero temperature, the Hellmann--Feynman theorem allows these derivatives to be evaluated as ground-state expectation values of the corresponding derivatives of the mean-field Hamiltonian in Eq.~\eqref{eq:hf_mf_hamiltonian} with respect to the mean-field parameters.
This gives
\begin{align}
    2P &= -\frac{\kondo}{N_0}\sum_{\bm{k},\sigma}
    \langle f_{\bm{k}\sigma}^{\dagger}c_{\bm{k}\sigma}\rangle,
    \notag\\
    1 &= \frac{1}{N_0}\sum_{\bm{k},\sigma}
    \langle f_{\bm{k}\sigma}^{\dagger}f_{\bm{k}\sigma}\rangle,
    \label{eq:kondo_saddle}
\end{align}
with respect to $P$ and $\lambda$.
Here and below, we set $k_{\mathrm B}=1$, and $N_0$ denotes the number of lattice unit cells.
The first equation in Eq.~\eqref{eq:kondo_saddle} determines the Kondo hybridization $P$, while the second equation imposes one $f$-fermion per site.
The fixed-density free energy of this phase is evaluated from the quasiparticle grand potential together with the mean-field constant terms and is compared with that of the magnetically ordered phase below.

\subsubsection{Antiferromagnetic Phase\label{sec:afm_method}}
In this subsection, we introduce the magnetic mean-field theory.
For an antiferromagnetic order with ordering wave vector $\bm{Q}$, the local-moment order parameter $\bm{M}$ generates an exchange field acting on the conduction electrons, and the conduction-electron polarization $\bm{m}$ gives a feedback field on the Abrikosov fermions through the Kondo exchange.
The magnetic order mixes electron states whose momenta differ by $\bm{Q}$, so the mean-field Hamiltonian is written in the folded Brillouin zone.
On the triangular lattice, we consider the 120-degree antiferromagnetic state favored by geometric frustration.
For the ordering wave vector $\bm{Q}=(4\pi/3,0)$, the folded Brillouin zone contains the three momentum sectors $\bm{k}$, $\bm{k}+\bm{Q}$, and $\bm{k}+2\bm{Q}$.
This magnetic reconstruction provides a useful comparison with the heavy-fermion phase, where a finite hybridization mixes the conduction-electron and local-moment degrees of freedom.
We use the spinors
\begin{align}
    \Psi_{\bm{k}}^{c} &=
    \left(
      c_{\bm{k}\uparrow},
      c_{\bm{k}\downarrow},
      c_{\bm{k}+\bm{Q}\uparrow},
      c_{\bm{k}+\bm{Q}\downarrow},
      c_{\bm{k}+2\bm{Q}\uparrow},
      c_{\bm{k}+2\bm{Q}\downarrow}
    \right)^T,
    \notag\\
    \Psi_{\bm{k}}^{f} &=
    \left(
      f_{\bm{k}\uparrow},
      f_{\bm{k}\downarrow},
      f_{\bm{k}+\bm{Q}\uparrow},
      f_{\bm{k}+\bm{Q}\downarrow},
      f_{\bm{k}+2\bm{Q}\uparrow},
      f_{\bm{k}+2\bm{Q}\downarrow}
    \right)^T.
    \label{eq:afm_spinors}
\end{align}
The complete AFM mean-field Hamiltonian can be written explicitly as
\begin{align}
    &H_{\mathrm{MF}}^{\mathrm{AFM}} \notag\\
    &=
    \sum_{\bm{k}}\sum_{a,b=0}^{2}\sum_{\sigma,\sigma'}
    c_{\bm{k}+a\bm{Q},\sigma}^{\dagger}
    [H_{cc}(\bm{k})]_{a\sigma,b\sigma'}
    c_{\bm{k}+b\bm{Q},\sigma'} \notag\\
    & +
    \sum_{\bm{k}}\sum_{a,b=0}^{2}\sum_{\sigma,\sigma'}
    f_{\bm{k}+a\bm{Q},\sigma}^{\dagger}
    [H_{ff}(\bm{k})]_{a\sigma,b\sigma'}
    f_{\bm{k}+b\bm{Q},\sigma'} \notag\\
    &
    +N_0\left(
      -\frac{z}{2}\heisen M_z^2
      +\frac{z}{4}\heisen M_\parallel^2
      -\kondo M_z m_z
      -\kondo M_\parallel m_\parallel
      -\lambda
    \right),
    \label{eq:afm_mf_hamiltonian}
\end{align}
Here $a,b$ label the three folded momentum sectors.
Using the combined spinor $\Psi_{\bm{k}}^{\mathrm{AFM}}=(\Psi_{\bm{k}}^{c},\Psi_{\bm{k}}^{f})^T$, the folded AFM Hamiltonian is
\begin{align}
    H_{\bm{k}}^{\mathrm{AFM}} =
    \begin{pmatrix}
        H_{cc}(\bm{k}) & 0 \\
        0 & H_{ff}(\bm{k})
    \end{pmatrix},
    \label{eq:Hk_afm}
\end{align}
with block matrices given by
\begin{widetext}
\begin{align}
H_{cc}(\bm{k})=
\begin{pmatrix}
\varepsilon_{\bm{k}}+\frac{\kondo}{2}M_z & 0 & 0 & \frac{\kondo}{2}M_\parallel & 0 & 0\\
0 & \varepsilon_{\bm{k}}-\frac{\kondo}{2}M_z & 0 & 0 & \frac{\kondo}{2}M_\parallel & 0\\
0 & 0 & \varepsilon_{\bm{k}+\bm{Q}}+\frac{\kondo}{2}M_z & 0 & 0 & \frac{\kondo}{2}M_\parallel\\
\frac{\kondo}{2}M_\parallel & 0 & 0 & \varepsilon_{\bm{k}+\bm{Q}}-\frac{\kondo}{2}M_z & 0 & 0\\
0 & \frac{\kondo}{2}M_\parallel & 0 & 0 & \varepsilon_{\bm{k}+2\bm{Q}}+\frac{\kondo}{2}M_z & 0\\
0 & 0 & \frac{\kondo}{2}M_\parallel & 0 & 0 & \varepsilon_{\bm{k}+2\bm{Q}}-\frac{\kondo}{2}M_z
\end{pmatrix},
\label{eq:Hcc}
\end{align}
\begin{align}
H_{ff}(\bm{k})=
\begin{pmatrix}
\lambda+\frac{1}{2}h_z & 0 & 0 & \frac{1}{2}h_\parallel & 0 & 0\\
0 & \lambda-\frac{1}{2}h_z & 0 & 0 & \frac{1}{2}h_\parallel  & 0\\
0 & 0 & \lambda+\frac{1}{2}h_z & 0 & 0 & \frac{1}{2}h_\parallel \\
\frac{1}{2}h_\parallel  & 0 & 0 & \lambda-\frac{1}{2}h_z & 0 & 0\\
0 & \frac{1}{2}h_\parallel  & 0 & 0 & \lambda+\frac{1}{2}h_z & 0\\
0 & 0 & \frac{1}{2}h_\parallel  & 0 & 0 & \lambda-\frac{1}{2}h_z
\end{pmatrix}.
\label{eq:Hff}
\end{align}
\end{widetext}

The matrices $H_{cc}$ and $H_{ff}$ act in the three folded momentum sectors $\bm{k}$, $\bm{k}+\bm{Q}$, and $\bm{k}+2\bm{Q}$.
Their dimension is $6\times6$ because each sector contains two spin states.
The variables $m_z$ and $m_\parallel$ are the out-of-plane and in-plane conduction-electron magnetizations.
The variables $M_z$ and $M_\parallel$ are the corresponding local-moment magnetizations.
The dispersions $\varepsilon_{\bm{k}+n\bm{Q}}$ with $n=0,1,2$ include the cavity self-energy in each folded sector.
The terms proportional to $M_z$ give a spin-dependent field within each sector.
The terms proportional to $M_\parallel$ connect opposite spins in neighboring sectors and describe the 120-degree spin rotation.
The $f$ fermions have the same structure with the effective fields $h_z$ and $h_\parallel$.
These fields are determined by $M$, $m$, $\heisen$, and $\kondo$ as
\begin{align}
    h_z=z\heisen M_z+\kondo m_z,
    \qquad
    h_\parallel=-\frac{z}{2}\heisen M_\parallel+\kondo m_\parallel.
    \label{eq:effective_fields}
\end{align}
Here $z=6$ is the coordination number of the triangular lattice.
The fixed-density mean-field energy of the AFM phase is
\begin{align}
    &E_{\mathrm{MF}}^{\mathrm{AFM}} \notag\\
    &=\frac{1}{3N_0}\sum_{\bm{k},\nu}
    E_{\bm{k}\nu}^{\mathrm{AFM}}f(E_{\bm{k}\nu}^{\mathrm{AFM}}) \notag\\
    &\quad
    -\frac{z}{2}\heisen M_z^2+\frac{z}{4}\heisen M_\parallel^2
    -\kondo M_z m_z - \kondo M_\parallel m_\parallel - \lambda
    +\mu\rho_{\mathrm e},
    \label{eq:energy_afm}
\end{align}
where $E_{\bm{k}\nu}^{\mathrm{AFM}}$ are the eigenvalues of Eq.~\eqref{eq:Hk_afm}.
The last five terms are the mean-field correction terms from the Heisenberg exchange, the Kondo exchange, and the constraint, and they are needed to avoid double counting of the exchange energy.
The self-consistent equations are obtained by setting the derivatives of the mean-field functional with respect to the mean fields to zero and can be written as
\begin{align}
    M_\parallel &=
    \frac{1}{3N_0}\sum_{\bm{k}}\sum_{n=0}^{2}
    \langle f_{\bm{k}+n\bm{Q},\uparrow}^{\dagger}
      f_{\bm{k}+(n+1)\bm{Q},\downarrow}\rangle,
    \notag\\
    m_\parallel &=
    \frac{1}{3N_0}\sum_{\bm{k}}\sum_{n=0}^{2}
    \langle c_{\bm{k}+n\bm{Q},\uparrow}^{\dagger}
      c_{\bm{k}+(n+1)\bm{Q},\downarrow}\rangle.
    \label{eq:afm_order_inplane}
\end{align}
\begin{align}
      M_z&=\frac{1}{6N_0}\sum_{\bm{k}}\sum_{n=0}^{2}
    \biggl(
      \langle f_{\bm{k}+n\bm{Q},\uparrow}^{\dagger}
      f_{\bm{k}+n\bm{Q},\uparrow}\rangle \notag\\
    &\hspace{35mm}  - \langle f_{\bm{k}+n\bm{Q},\downarrow}^{\dagger}
      f_{\bm{k}+n\bm{Q},\downarrow}\rangle
    \biggr),
    \notag\\
    m_z&=\frac{1}{6N_0}\sum_{\bm{k}}\sum_{n=0}^{2}
    \biggl(
      \langle c_{\bm{k}+n\bm{Q},\uparrow}^{\dagger}
      c_{\bm{k}+n\bm{Q},\uparrow}\rangle\notag\\
    &\hspace{35mm}  - \langle c_{\bm{k}+n\bm{Q},\downarrow}^{\dagger}
      c_{\bm{k}+n\bm{Q},\downarrow}\rangle
    \biggr),
    \label{eq:afm_order_z}
\end{align}
\begin{align}
    1 = \frac{1}{3N_0}\sum_{\bm{k}}\sum_{n=0}^{2}\sum_{\sigma}
    \langle
      f_{\bm{k}+n\bm{Q},\sigma}^{\dagger}f_{\bm{k}+n\bm{Q},\sigma}
    \rangle.
    \label{eq:afm_constraint}
\end{align}
The last equation imposes the constraint on the average number of $f$-fermions in the AFM phase.

We also examined the coexistence phase with both Kondo hybridization and 120-degree magnetic order in representative parameter ranges by adding the off-diagonal hybridization term $P$ to Eq.~\eqref{eq:Hk_afm}, in the same way as in the heavy-fermion phase~\cite{liPhaseEvolutionTwodimensional2015}.
In the numerical calculations, we did not find this phase as the lowest-energy solution in the parameter range examined here.

\begin{figure}[t]
  \centering
  \includegraphics[width=\columnwidth]{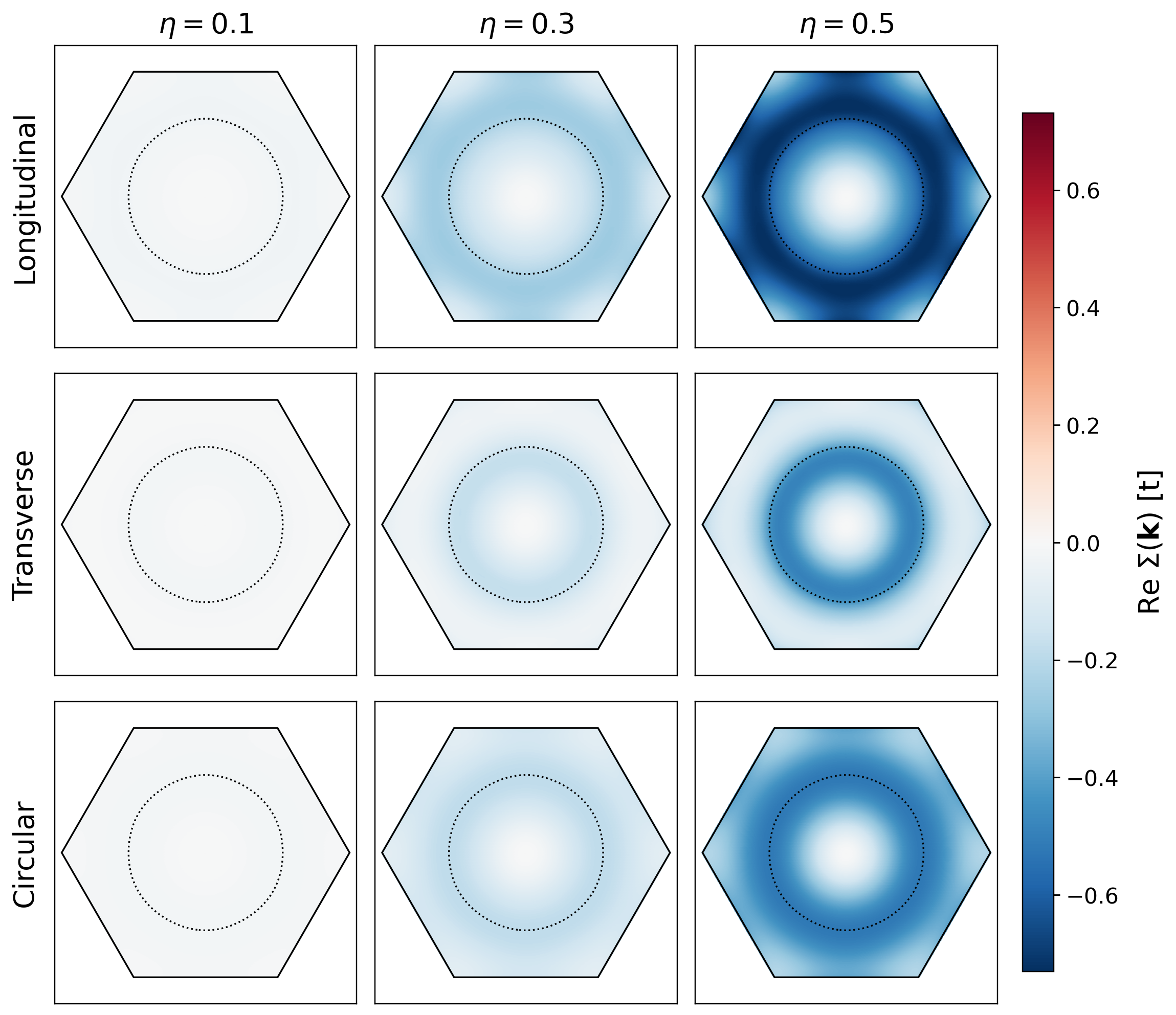}
  \caption{The self-energy $\Sigma(\bm{k})$ in momentum space for the triangular lattice at $\rho_{\mathrm e}=0.7$.
  The black dotted line shows the bare conduction-electron Fermi surface at this filling.
  The self-energy has a clear momentum dependence, and the pattern changes with the polarization structure.
  From top to bottom, these are the longitudinal structure and the transverse and circular model structures, respectively.}
  \label{fig:sigma_bz}
\end{figure}

\subsubsection{Ferromagnetic Phase\label{sec:fm_method}}
To examine the effect of the cavity self-energy at low electron density, we also consider a ferromagnetic phase as a model comparison.
We assume a fixed interaction $\heisen<0$ and ferromagnetic order in which all local moments point along the same spin direction.
We define the local-moment magnetization $M$ and the conduction-electron magnetization $m$ as
\begin{align}
    M &= \frac{1}{2}
      \left\langle
        f_{i\uparrow}^{\dagger}f_{i\uparrow}
        -f_{i\downarrow}^{\dagger}f_{i\downarrow}
      \right\rangle,
    \notag\\
    m &= \frac{1}{2}
      \left\langle
        c_{i\uparrow}^{\dagger}c_{i\uparrow}
        -c_{i\downarrow}^{\dagger}c_{i\downarrow}
      \right\rangle.
    \label{eq:fm_order_parameters}
\end{align}
The ferromagnetic mean-field Hamiltonian is
\begin{align}
    H_{\mathrm{MF}}^{\mathrm{FM}}
    &= \sum_{\bm{k},\sigma=\pm1}
      \left(
        \varepsilon_{\bm{k}}+\frac{\sigma\kondo M}{2}
      \right)
      c_{\bm{k}\sigma}^{\dagger}c_{\bm{k}\sigma}
    \notag\\
    &\quad+ \sum_{\bm{k},\sigma=\pm1}
      \left[
        \lambda+\frac{\sigma}{2} \left(z\heisen M+\kondo m\right)
      \right]
      f_{\bm{k}\sigma}^{\dagger}f_{\bm{k}\sigma}
    \notag\\
    &\quad + N_0 \left(
      -\frac{z\heisen M^2}{2}-\kondo Mm-\lambda
    \right).
    \label{eq:fm_mf_hamiltonian}
\end{align}
Here $z=6$ is the coordination number of the triangular lattice.
The fields proportional to $\kondo M$ and $z\heisen M+\kondo m$ split the conduction-electron and $f$-fermion bands, respectively.
The parameters $M$, $m$, $\lambda$, and $\mu$ are determined self-consistently at fixed conduction-electron filling.
In the restricted ferromagnetic phase considered here, the Kondo hybridization is set to zero.
The low-density calculation below is restricted to a comparison between the pure FM and heavy-fermion phases.

\subsection{Numerical Implementation}

For concrete calculations, we use a Gaussian distribution for the coupling factor appearing in Eq.~\eqref{eq:self_energy_intro}.
\begin{align}
    \frac{g_{\bm{q}}^2}{\omega_{\bm{q}}^2} =
    \frac{\eta^2}{D_0}\exp\left[-\frac{|\bm{q}-\bm{q}^\ast|^2}{2q_c^2}\right].
    \label{eq:eta_distribution}
\end{align}
Here $D_0=\sum_{\bm q}\exp[-|\bm q-\bm q^\ast|^2/(2q_c^2)]$ is a normalized factor for $\eta$.
For the nearest-neighbor triangular lattice used in the numerical calculations, the bare conduction-band dispersion is
$\varepsilon^0_{\bm{k}}=-2t[\cos(k_xa)+2\cos(k_xa/2)\cos(\sqrt{3}k_ya/2)]$.
For $t>0$, its bandwidth is $W=9t$.
Throughout the numerical calculations, we set the hopping amplitude and lattice constant to $t=1$ and $a=1$.
Here we do not specify $g_{\bm{q}}$ and $\omega_{\bm{q}}$ as independent inputs, but directly parametrize the momentum dependence of $g_{\bm{q}}^2/\omega_{\bm{q}}^2$ by Eq.~\eqref{eq:eta_distribution}.
For the h-BN hyperbolic phonon-polariton modes motivating the longitudinal case, large in-plane momenta are supported while the mode energies remain within a bounded spectral window of approximately $155\text{--}200\,\mathrm{meV}$ ~\cite{caldwellSubdiffractionalVolumeconfinedPolaritons2014}.
Thus, these modes can mediate a broad range of in-plane momentum transfers without corresponding increases in their energies.
In the present minimal model, the resulting momentum dependences of the frequency, mode normalization, and coupling are represented effectively through the Gaussian profile of $g_{\bm{q}}^2/\omega_{\bm{q}}^2$.
Related cavity-moir\'e calculations found that including the finite momentum width of $g_{\bm q}^2/\omega_{\bm q}^2$ only weakly modifies the resulting phase diagram and leaves its qualitative structure unchanged; thus we use a minimal effective profile in the present qualitative analysis~\cite{masukiCavityMoireMaterials2024}.

We use $\bm{q}^\ast=0$ and $q_c=1$ in the reciprocal-lattice units of the triangular lattice.
For the metallic comparison between the AFM and heavy-fermion phases at $\rho_{\mathrm e}=0.7$, we fix $\heisen=0.15t$ and vary $\kondo$ and $\eta$.
For the low-density comparison between the FM and heavy-fermion phases, we use $\rho_{\mathrm e}=0.4$ and a fixed interaction $\heisen=-0.15t$.

We compare three in-plane polarization structures.
The polarization vectors are defined as
\begin{align}
    \bm{e}_{\bm{q}}^{\mathrm{L}}&=\frac{\bm{q}}{|\bm{q}|},\\
    \bm{e}_{\bm{q}}^{\mathrm{T}}&=\frac{\hat{\bm{e}}_z\times\bm{q}}{|\bm{q}|},\\
    \bm{e}_{\bm{q}}^{\mathrm{C}}&=\frac{1}{\sqrt{2}}\left(\frac{\bm{q}}{|\bm{q}|} + i\frac{\hat{\bm{e}}_z\times\bm{q}}{|\bm{q}|}\right),
    \label{eq:polarizations}
\end{align}
for the longitudinal, transverse, and circular in-plane polarizations, respectively.
The longitudinal projected field is the physical case motivated by high-momentum hyperbolic phonon polaritons in h-BN.
Actual h-BN modes can have an in-plane transverse component, but this component is very small in the large-$|\bm{q}|$ region that is important for the self-energy.
The transverse and circular structures are model cases used for comparison and are not equally realistic descriptions of the h-BN modes.
The full three-dimensional field remains transverse in the Coulomb gauge even when its projection onto the two-dimensional material plane is longitudinal.
More generally, evanescent near fields at two-dimensional interfaces can acquire elliptical polarization because spatial gradients and anisotropic mode profiles generate phase-shifted field components.
For a TMD layer placed close to h-BN, the evanescent field decays away from the interface over a length scale of order $1/|\bm{q}|$.
The circular in-plane structure $\bm{e}_{\bm{q}}^{\mathrm{C}}$ can therefore be regarded as an idealized model of the polarization textures that may arise in such spatially varying near fields.

\section{Results\label{sec:results}}
\begin{figure}[tbp]
  \centering
  \includegraphics[width=0.78\columnwidth]{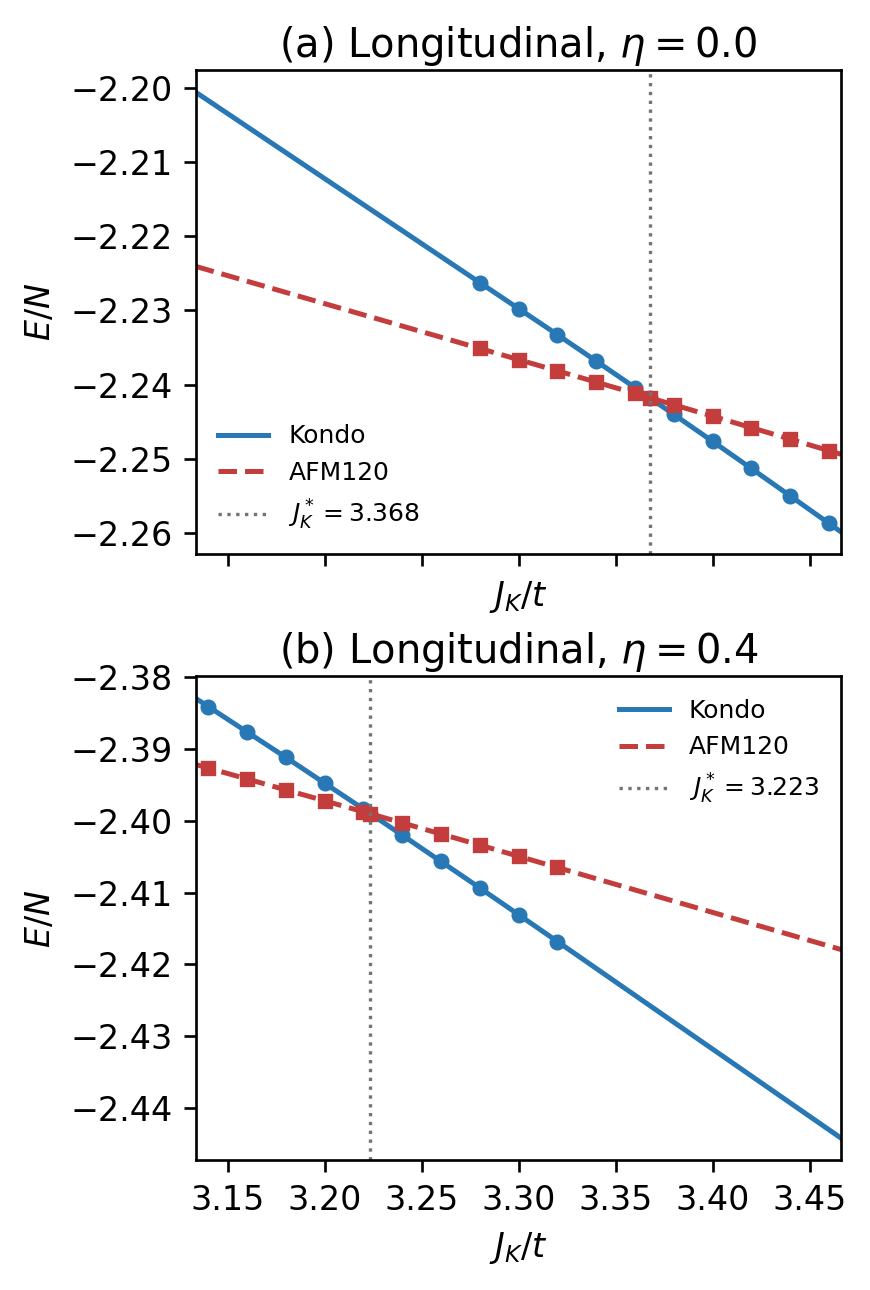}
  \caption{Energy crossing between the heavy-fermion phase and the 120-degree antiferromagnetic phase at $\rho_{\mathrm e}=0.7$ for the longitudinal structure.
  The upper and lower panels show $\eta=0$ and $\eta=0.4$, respectively.
  The two panels use the same horizontal range.
  The 120-degree AFM phase has the lower energy at smaller $\kondo$, while the heavy-fermion phase becomes lower at larger $\kondo$.}
  \label{fig:kafm_crossing}
\end{figure}

\subsection{Cavity-Induced Self-Energy}
We first examine how the cavity-induced self-energy changes the low-energy conduction-electron states.
The self-energy shows a clear momentum dependence, and its pattern depends on the polarization as shown in Fig.~\ref{fig:sigma_bz}.
The dotted line in this figure shows the bare conduction-electron Fermi surface at $\rho_{\mathrm e}=0.7$.
Because the first-order self-energy is linear in the effective interaction $g_{\bm{q}}^2/\omega_{\bm{q}}^2 \propto \eta^2$ in Eq.~\eqref{eq:eta_distribution}, the characteristic scale of $\Sigma(\bm{k})$ grows as $O(\eta^2)$.
Numerically, at $\eta\simeq0.5$ the largest longitudinal correction reaches $\max_{\bm{k}}|\Sigma(\bm{k})|\simeq0.74t$, which is of the same order as the bare hopping scale.
This order-of-magnitude estimate shows that the cavity-mediated interaction can produce a dispersion correction comparable to the microscopic one-particle scale in the displayed parameter range.
Since the ordinary Coulomb-induced renormalization is treated as part of the reference band, the calculation should be read as the additional cavity-induced shift relative to that effective band.
This estimate also indicates that the most controlled side of the perturbative treatment is the smaller-$\eta$ regime.

We recall that the Kondo effect is sensitive to the density of states at the Fermi level.
In a simple impurity estimate, the Kondo temperature is written as
\begin{align}
    T_{\mathrm{K}}= D\exp\left[-\frac{1}{\kondo\rho(\varepsilon_F)}\right],
    \label{eq:tk_estimate}
\end{align}
where $D$ is the bandwidth and $\rho(\varepsilon_F)$ is the conduction-electron density of states at the Fermi energy.

For a single magnetic impurity, a similar cavity-induced interaction enhances the Kondo temperature through an effective mass enhancement of the conduction electron~\cite{mochidaCavityenhancedKondoEffect2024}.
In the present lattice problem, this mass-enhancement intuition remains useful for the longitudinal projected-field type, while the same self-energy also changes the magnetic phase energy through the conduction-electron band.
Thus, the transition point is determined by the relative energy changes of the heavy-fermion and AFM phases, as discussed in Sec.~\ref{sec:discussion}.

\subsection{Modified Doniach Phase Transition}

\begin{figure}[tbp]
  \centering
  \includegraphics[width=\columnwidth]{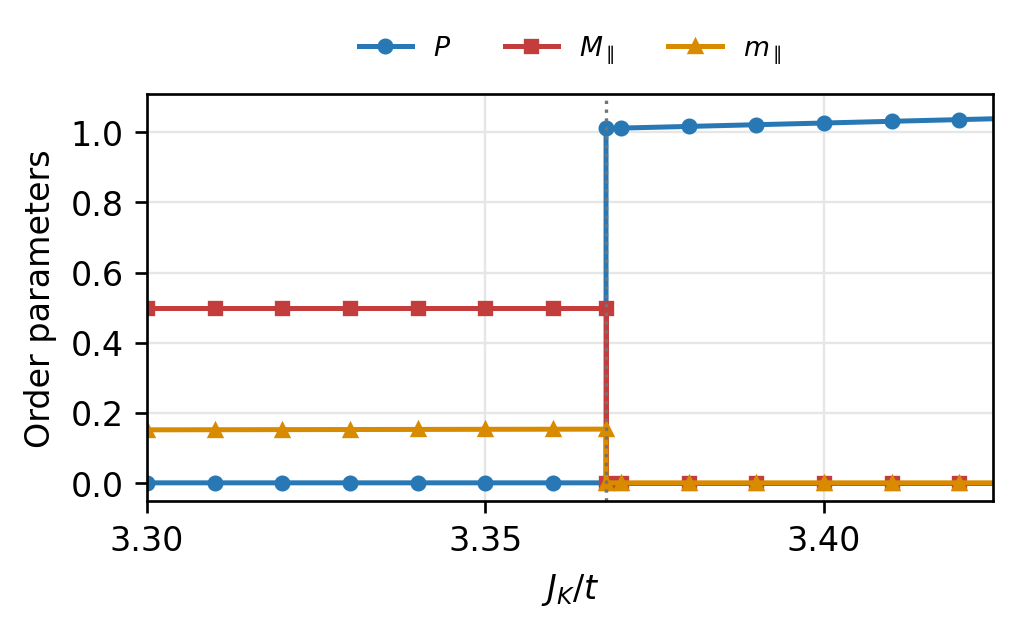}
  \caption{The mean-field parameters of the ground states near the phase transition point at $\rho_{\mathrm e}=0.7$ and $\eta=0$.
  The Kondo hybridization $P$ is finite in the heavy-fermion phase, while the in-plane magnetic order parameter $M_\parallel$ characterizes the 120-degree AFM phase.}
  \label{fig:mfparam}
\end{figure}

\begin{figure*}[!t]
  \centering
  \includegraphics[width=\textwidth]{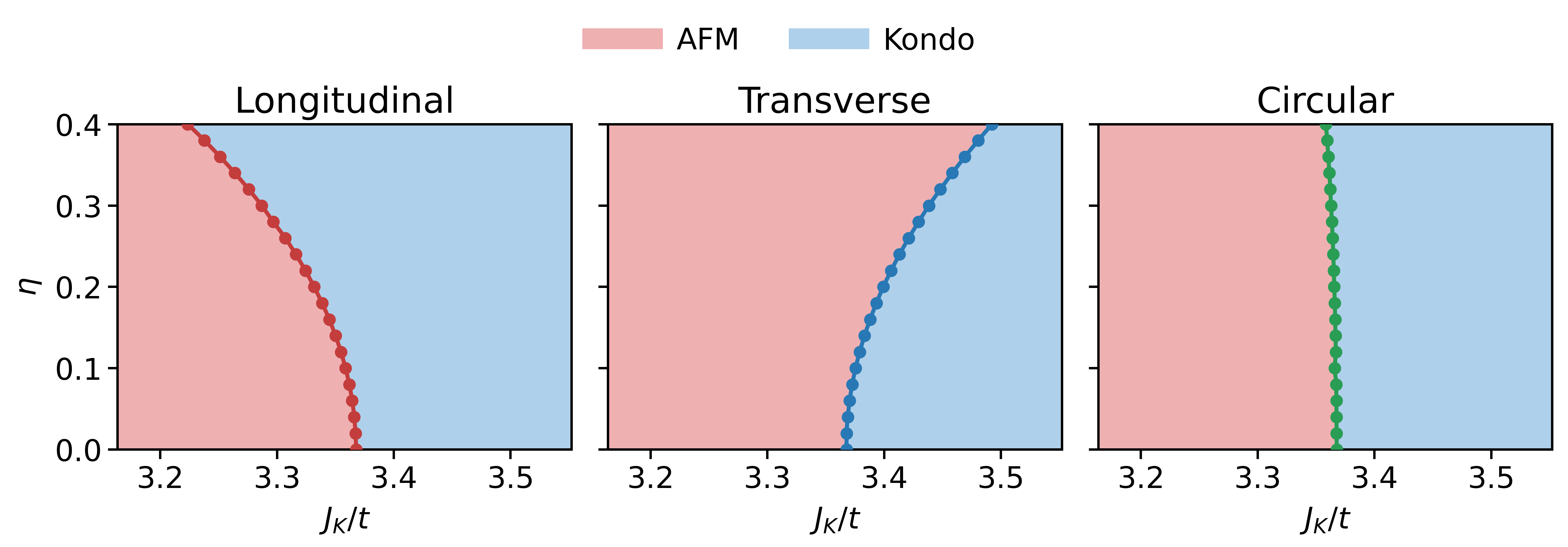}
  \caption{$J_{\mathrm K}$--$\eta$ phase diagram for the triangular lattice at $\rho_{\mathrm e}=0.7$.
  The phase diagram is obtained by solving the self-consistent equations for each phase and comparing the corresponding ground-state energies.
  The horizontal axis shows the Kondo coupling $J_{\mathrm K}$, and the vertical axis shows the cavity coupling strength $\eta$.
  The heavy-fermion region expands toward smaller $\kondo$ for the longitudinal case, while the 120-degree AFM region expands for the transverse model case.
  The circular model case gives a transition line between these two cases.}
  \label{fig:kafm_phase}
\end{figure*}

\begin{figure}[!t]
  \centering
  \includegraphics[width=\columnwidth]{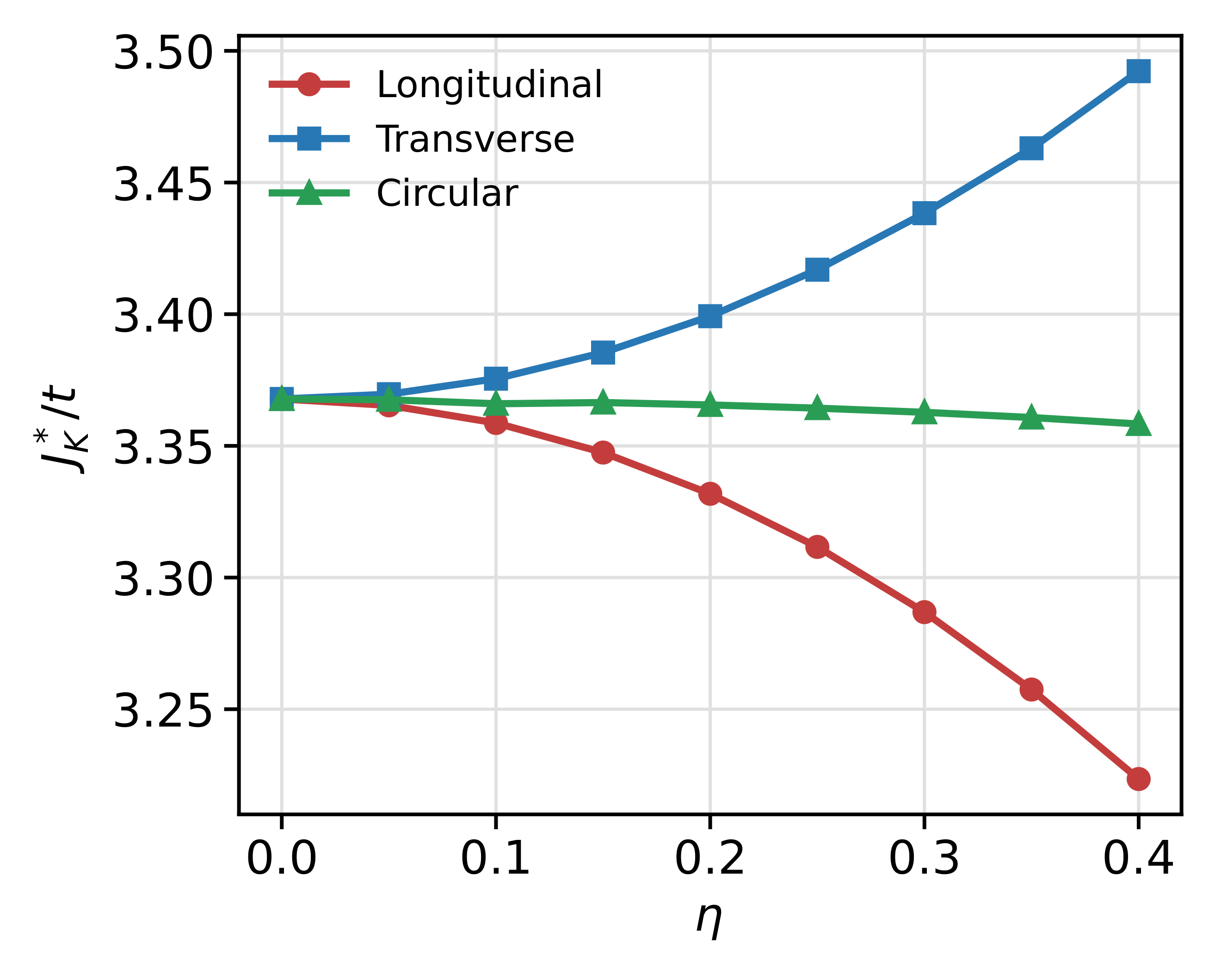}
  \caption{Extracted transition coupling $\kondo^\ast$ at $\rho_{\mathrm e}=0.7$ as a function of the cavity strength $\eta$ for all polarization structures. The longitudinal case shifts the transition to smaller $\kondo^\ast$. The transverse model case shifts the transition to larger $\kondo^\ast$, and the circular model case lies between these two shifts.}
  \label{fig:kafm_transition_eta}
\end{figure}

\begin{figure*}[!t]
  \centering
  \includegraphics[width=\textwidth]{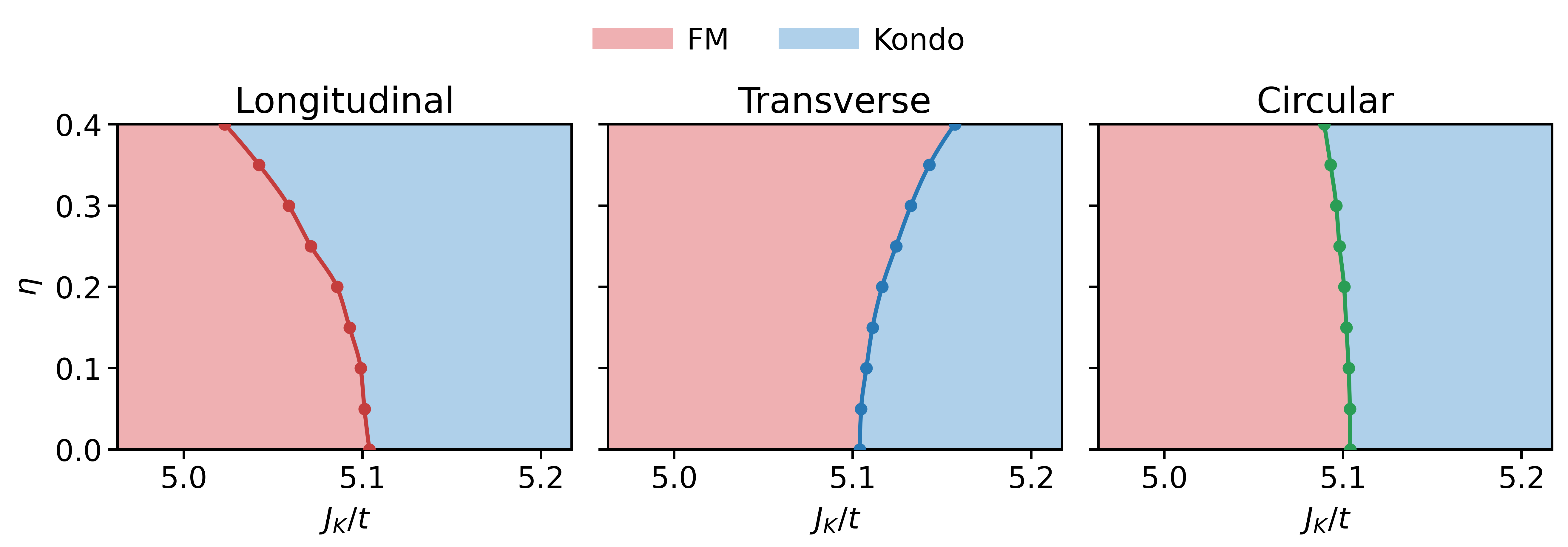}
  \caption{$J_{\mathrm{K}}$--$\eta$ phase diagram at $\rho_{\mathrm e}=0.4$ for a fixed ferromagnetic interaction $J_{\mathrm H}=-0.15t$.
  The phase boundary separates the ferromagnetic and heavy-fermion phases.
  The heavy-fermion region expands toward smaller $\kondo$ for the longitudinal case, while the FM region expands for the transverse model case.
  The circular model case gives a transition line between these two cases.}
  \label{fig:rho04_fm_transition_eta}
\end{figure*}

We next study the quantum phase transition between the heavy-fermion phase and the 120-degree AFM phase on a triangular lattice.
For each value of $\eta$ and for each polarization, we solve the two mean-field problems separately and compare their converged energies.
The transition point $\kondo^\ast$ is determined by the crossing of these two energies.
In the following, we present the results at the metallic filling $\rho_{\mathrm e}=0.7$.
At this filling, the transition can be discussed together with a reconstruction of the Fermi-surface volume.
The corresponding results for the square lattice, the filling dependence, and the half-filled triangular lattice are presented in Appendixes~\ref{app:square} and \ref{app:filling}.

Figure~\ref{fig:kafm_crossing} shows an energy crossing.
On the small-$\kondo$ side, the 120-degree AFM solution has the lower energy.
On the large-$\kondo$ side, the heavy-fermion solution becomes lower in energy.
The corresponding mean-field parameters are shown in Fig.~\ref{fig:mfparam}.
Across the transition, the Kondo hybridization $P$ becomes the dominant order parameter in the heavy-fermion phase, while the in-plane magnetic order parameter $M_\parallel$ characterizes the 120-degree AFM phase.
This behavior is consistent with a description of the competition between Kondo screening and magnetic ordering in the standard Doniach-type quantum phase transition.
As shown in Fig.~\ref{fig:mfparam}, the order parameters in the ground state switch before and after the transition point.

In the heavy-fermion phase, the exchange-energy gain from hybridization competes with an increase in the conduction-electron kinetic energy.
The magnetic phase is stabilized both by the Heisenberg interaction and by the conduction-electron polarization induced by the Kondo exchange.
For the fixed $\heisen=0.15t$ used here, this competition places the zero-cavity crossing at $\kondo^\ast(0)\simeq3.37t$, above which the heavy-fermion state has the lower energy.
With the bandwidth $W=9t$, the present crossing corresponds to $\kondo^\ast(0)/W\simeq0.37$, placing the Kondo exchange in an intermediate-coupling regime.
The precise crossing depends on the filling, band dispersion, magnetic interaction, and mean-field treatment.
For comparison, the fermionic mean-field study of the half-filled square-lattice Kondo model by Li et al.~\cite{liPhaseEvolutionTwodimensional2015} found that the pure antiferromagnetic phase gives way to a coexistence phase at $\kondo\simeq3.15t$ and that the Kondo paramagnetic phase sets in at $\kondo\simeq3.42t$.
Gate-tunable moir\'e Kondo-lattice systems provide promising platforms for exploring this competition over a tunable range of coupling strengths~\cite{kumarGatetunableHeavyFermion2022,guerciChiralKondoLattice2023,zhaoGatetunableHeavyFermions2023}.
For example, the moir\'e model of Kumar et al.~\cite{kumarGatetunableHeavyFermion2022} uses $\kondo\simeq22\,\mathrm{meV}$ in a representative calculation with a conduction-band width $W\simeq104\,\mathrm{meV}$, giving $\kondo/W\simeq0.2$.

The extracted transition points are summarized in Figs.~\ref{fig:kafm_phase} and \ref{fig:kafm_transition_eta}.
The transition line shows a clear dependence on the in-plane polarization structure.
For the longitudinal case, the transition shifts to smaller $\kondo$, which means that the heavy-fermion phase is stabilized relative to the 120-degree AFM phase.
For the transverse model case, the transition shifts to larger $\kondo$, which means that the 120-degree AFM phase is relatively stabilized.
The circular model case lies between the longitudinal and transverse cases.
These results indicate that the cavity changes the competition between the two phases through the momentum structure of $\Sigma(\bm{k})$.

\subsection{Low-Density Ferromagnetic Comparison\label{sec:fm_results}}
We next consider a low-density case at $\rho_{\mathrm e}=0.4$, where we assume a fixed ferromagnetic interaction $\heisen=-0.15t$.
The purpose of this calculation is not to derive the sign of the magnetic interaction from the conduction-electron susceptibility.
Instead, we examine how the cavity-induced self-energy changes the competition between the heavy-fermion phase and a uniform ferromagnetic phase.

Figure~\ref{fig:rho04_fm_transition_eta} shows the resulting $\kondo$--$\eta$ phase boundaries.
The longitudinal structure shifts the transition toward the heavy-fermion side, whereas the transverse structure stabilizes the FM phase.
The circular structure produces a smaller shift between these two limits.
The ordering of the polarization effects is the same as in the AFM--heavy-fermion comparison.

\begin{figure*}[!t]
  \centering
  \includegraphics[width=\textwidth]{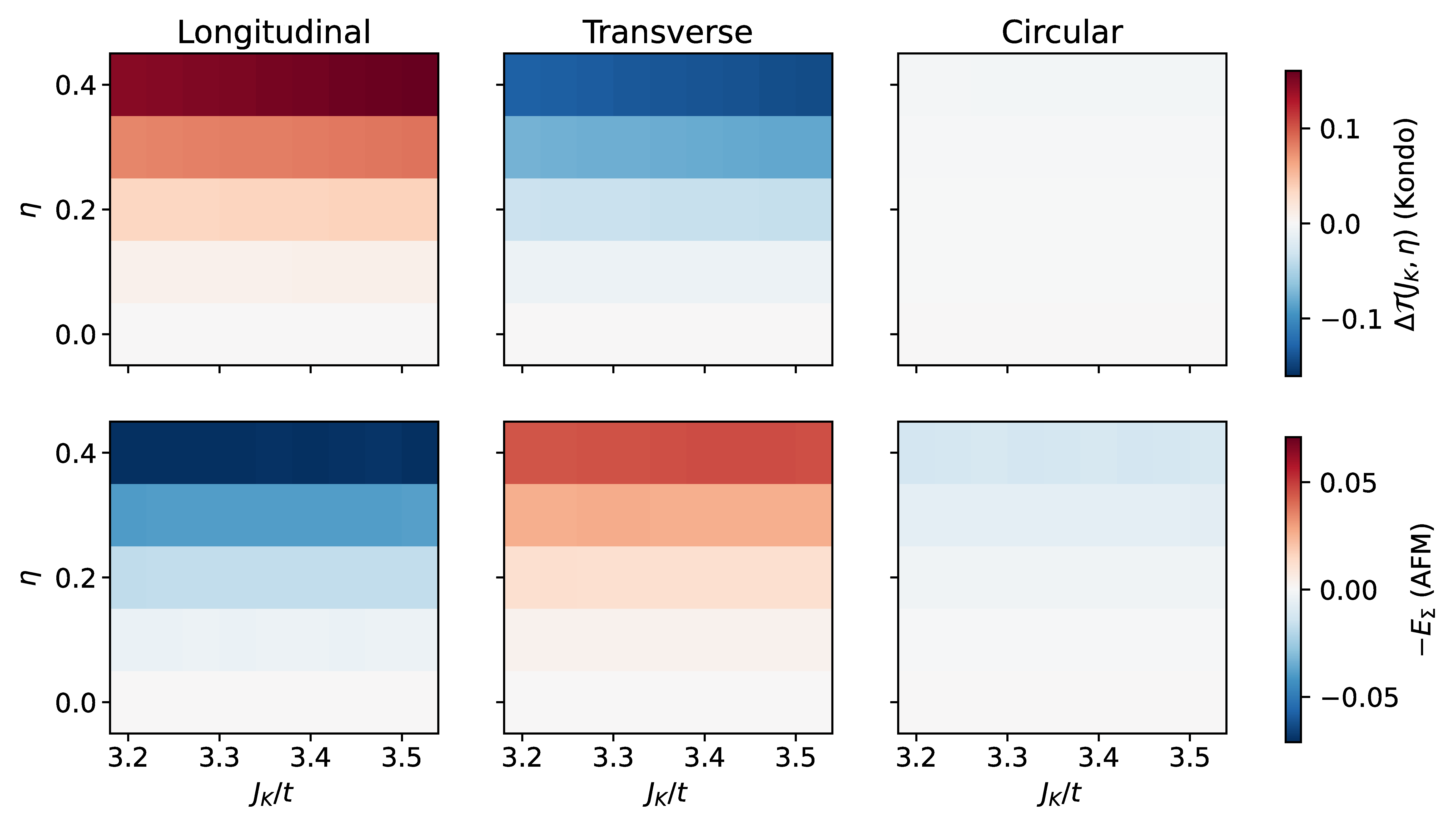}
  \caption{Two numerical measures of how the cavity self-energy changes the competing phase energies at $\rho_{\mathrm e}=0.7$.
  The top row shows the cavity-induced change $\Delta\mathcal{T}(\kondo,\eta)=\mathcal{T}(\kondo,\eta)-\mathcal{T}(\kondo,0)$ obtained from the $P\rightarrow0$ self-consistent equation.
  Positive values mean that the cavity enhances the tendency toward infinitesimal Kondo hybridization relative to $\eta=0$.
  The bottom row shows the antiferromagnetic self-energy contribution $-E_{\Sigma}^{\mathrm{AFM}}$ obtained from the nonuniform self-energy in the conduction sector of the self-consistent 120-degree antiferromagnetic quasiparticles.
  Positive values mean that the folded antiferromagnetic conduction-electron distribution gives an energy gain from the nonuniform self-energy.}
  \label{fig:phase_filter_maps}
\end{figure*}

\begin{figure}[!t]
  \centering
  \includegraphics[width=\columnwidth]{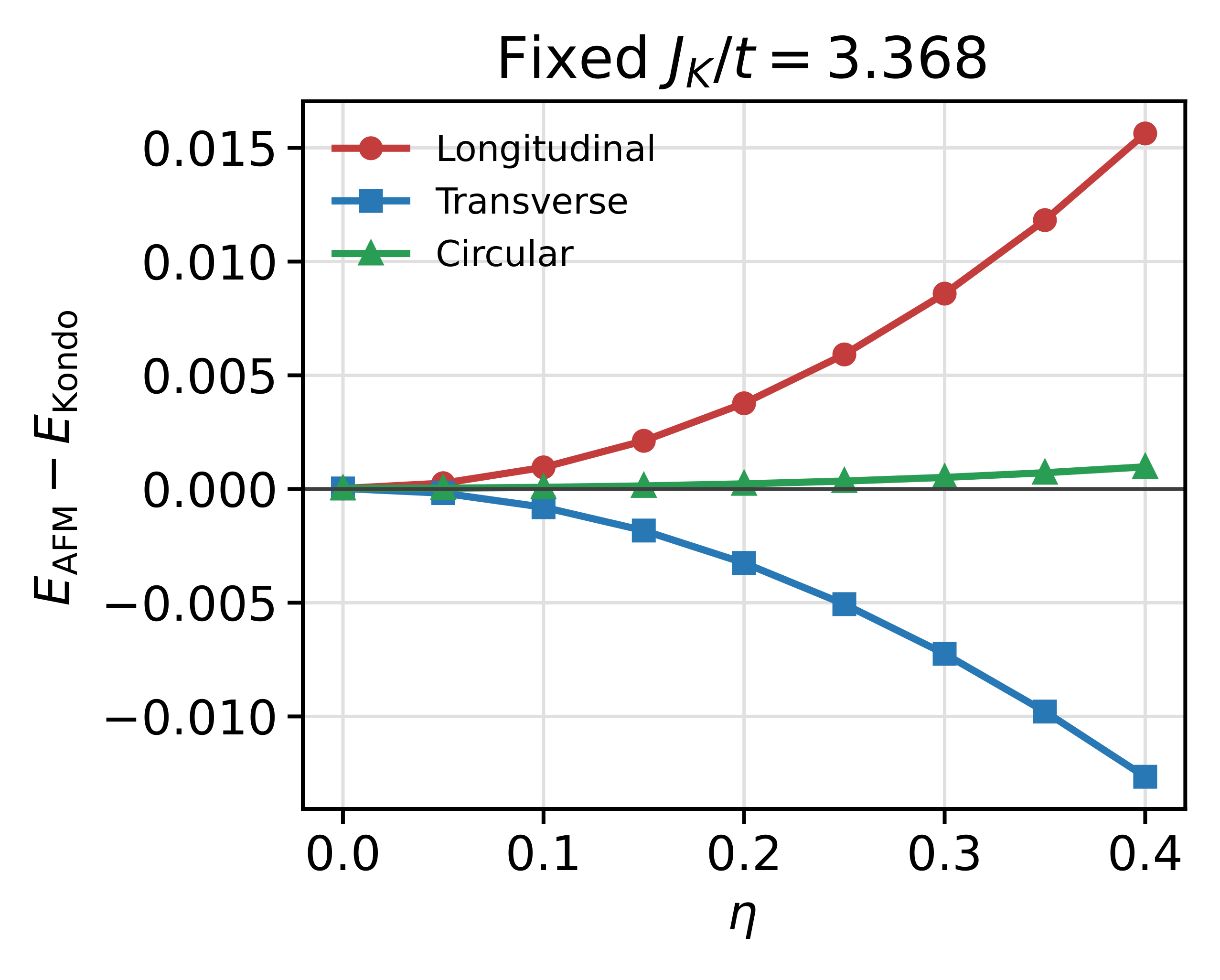}
  \caption{Relative self-consistent phase energy at $\rho_{\mathrm e}=0.7$ and fixed $\kondo/t\simeq3.37$, corresponding to the zero-cavity transition point.
  The plotted quantity is $\delta E=E_{\mathrm{AFM}}-E_{\mathrm{Kondo}}$.
  Positive values favor the heavy-fermion phase, while negative values favor the 120-degree antiferromagnetic phase.}
  \label{fig:phase_energy_difference}
\end{figure}

\section{Discussion\label{sec:discussion}}
We give a physical interpretation of the polarization-dependent shift of the phase boundary.
The cavity enters the Kondo lattice mainly through the Fock self-energy of the conduction electrons.
This self-energy first lowers the conduction-electron energy in a momentum-dependent way, and the heavy-fermion and antiferromagnetic mean-field states then convert these energy modifications into the transition point shift.

\subsection{Self-Energy Influence}
The qualitative structure of Fig.~\ref{fig:sigma_bz} follows directly from the polarization factor in Eq.~\eqref{eq:self_energy_intro}.
In the discrete notation, the same expression can be read as
\begin{align}
    \Sigma_\alpha(\bm{k})
    =
    -2\sum_{\bm{q}}
    \frac{\hbar^2g_{\bm{q}}^2}{m\omega_{\bm{q}}^2}
    \bigl|\bm{e}_{\bm{q}}^\alpha\cdot\bm{k}\bigr|^2
    f(\varepsilon_{\bm{k}-\bm{q}}),
    \label{eq:discussion_sigma_q}
\end{align}
where $\alpha=\mathrm{L},\mathrm{T},\mathrm{C}$.
This self-energy is negative and means that a conduction-electron state at momentum $\bm{k}$ is lowered when it is connected to occupied states $\bm{k}-\bm{q}$ by virtual cavity-mediated scattering with a large polarization factor.
The factor $f(\varepsilon_{\bm{k}-\bm{q}})$ selects the occupied state reached by subtracting the virtual momentum transfer $\bm{q}$ from $\bm{k}$.
Thus Eq.~\eqref{eq:discussion_sigma_q} says that the energy of a state at $\bm{k}$ is lowered by occupied states at $\bm{k}-\bm{q}$, with a polarization-dependent weight for that virtual scattering path.

For a longitudinal projected field, the factor $|\bm{e}_{\bm{q}}\cdot\bm{k}|^2$ weights the component of $\bm{k}$ parallel to the virtual momentum transfer $\bm{q}$.
For a transverse in-plane structure, it weights the orthogonal component.
Equivalently,
\begin{align}
    \bigl|\bm{e}_{\bm{q}}^{\mathrm{L}}\cdot\bm{k}\bigr|^2
    &= k^2\cos^2(\phi_q-\phi_k),
    \\
    \bigl|\bm{e}_{\bm{q}}^{\mathrm{T}}\cdot\bm{k}\bigr|^2
    &= k^2\sin^2(\phi_q-\phi_k).
    \label{eq:discussion_polarization_weights}
\end{align}
The Fermi factor $f(\varepsilon_{\bm{k}-\bm{q}})$ in Eq.~\eqref{eq:self_energy_intro} then restricts this angular weight to momentum transfers connected to occupied states.
Thus, the longitudinal case counts virtual scattering paths in which $\bm{k}$ is nearly parallel to $\bm{q}$, while the transverse case counts paths in which $\bm{k}$ is nearly perpendicular to $\bm{q}$.
This difference gives a broad radial deformation for the longitudinal case, whose largest negative values appear toward the Brillouin-zone edge.
The transverse model structure instead gives a more annular pattern around the bare Fermi surface, because it emphasizes tangential virtual scattering around the occupied Fermi sea.
The circular structure averages the two angular channels and therefore produces a pattern between these two limits.

\subsection{Energy Changes in the Competing Phases\label{sec:energy_response}}
The phase-boundary shift is then determined by how this momentum-dependent change of the conduction band enters the competing phase energies.
The uniform part of $\Sigma(\bm{k})$ is mostly absorbed into the chemical potential at fixed $\rho_{\mathrm e}$.
The remaining momentum dependence matters because it lowers some conduction-electron states more than others.
At first order, this is simply the expectation value of the self-energy term in the corresponding mean-field state.
Therefore, a phase gains more energy when its conduction-electron component occupies, or hybridizes through, the momenta where the nonuniform self-energy is more negative.

For the heavy-fermion phase, the relevant momentum-space distribution is the conduction-electron component of the hybridized $c$-$f$ quasiparticles.
The self-energy enters the heavy-fermion energy in Eq.~\eqref{eq:energy_hf} through the quasiparticle energies $E_{\bm{k}}^\nu$.
It changes $E_{\bm{k}}^\nu$ because the conduction level inside the $c$-$f$ hybridized Hamiltonian is $\varepsilon_{\bm{k}}=\varepsilon^0_{\bm{k}}+\Sigma(\bm{k})-\mu$.
The optimized heavy-fermion energy correction is therefore most favorable when the momentum-dependent self-energy acts on conduction-electron states that strongly participate in the hybridized quasiparticles.

For the magnetic phases, the relevant momentum-space distributions are the conduction-electron components of the corresponding magnetic quasiparticles.
Unlike the heavy-fermion Hamiltonian, where $P$ directly hybridizes the $c$ and $f$ fermions, the magnetic Hamiltonians are block diagonal in the $c$- and $f$-fermion sectors.
These sectors are denoted by $H_{cc}$ and $H_{ff}$ in the AFM phase.
The Kondo exchange still relates the two sectors self-consistently because the $f$-fermion magnetization $M$ enters the conduction-electron block, while the conduction-electron magnetization $m$ enters the $f$-fermion block.
In the FM phase, Eq.~\eqref{eq:fm_mf_hamiltonian} shows that the self-energy enters the spin-resolved conduction-electron dispersions $\varepsilon_{\bm{k}}+\sigma\kondo M/2$.
In the 120-degree AFM phase, Eq.~\eqref{eq:afm_mf_hamiltonian} shows that it enters the conduction block $H_{cc}(\bm{k})$ through $\varepsilon_{\bm{k}}$, $\varepsilon_{\bm{k}+\bm{Q}}$, and $\varepsilon_{\bm{k}+2\bm{Q}}$.
After diagonalization, the FM and AFM phases sample the same nonuniform self-energy through different conduction-electron distributions.
The phase boundaries in Figs.~\ref{fig:kafm_phase} and \ref{fig:rho04_fm_transition_eta} follow a common polarization trend.
The longitudinal structure favors the heavy-fermion phase, the transverse structure favors the magnetic phases, and the circular structure lies between these two cases.

To clarify the origin of the phase-boundary shift, we introduce complementary diagnostics for the heavy-fermion and antiferromagnetic phases.
On the heavy-fermion side, the $P\rightarrow0$ limit of the self-consistent equation gives the onset condition for Kondo hybridization~\cite{sachdevQuantumPhasesMatter2023}.
Subtracting unity from the right-hand side of this condition, we define
\begin{align}
    \mathcal{T}(\kondo,\eta) =\frac{\kondo}{N_0}\sum_{\bm{k}}\frac{f(\varepsilon_{\bm{k}})-f(\lambda)}{\lambda-\varepsilon_{\bm{k}}}-1.
    \label{eq:kondo_pzero}
\end{align}
The $P\to0$ self-consistent condition is therefore $\mathcal{T}(\kondo ,\eta)=0$.
A value $\mathcal{T}(\kondo ,\eta)>0$ means that the right-hand side exceeds unity and favors an infinitesimal Kondo hybridization.
To isolate the cavity-induced response, the top row of Fig.~\ref{fig:phase_filter_maps} shows $\Delta\mathcal{T}(\kondo,\eta)=\mathcal{T}(\kondo,\eta)-\mathcal{T}(\kondo,0)$.
On the antiferromagnetic side it can be obtained by weighting the nonuniform self-energy $\delta\Sigma(\bm{k})=\Sigma(\bm{k})-\Sigma_0$, where $\Sigma_0$ is a mean value of the self-energy in the folded conduction-electron sectors of the self-consistent 120-degree antiferromagnetic quasiparticles.
After rebuilding the 120-degree antiferromagnetic Hamiltonian with the self-consistent order parameters, we evaluate
\begin{align}
  E_{\Sigma}^{\mathrm{AFM}}
    = \frac{1}{3N_k} \sum_{\bm{k},b} f(E_{\bm{k}b}^{\mathrm{AFM}})\bra{\bm{k},b}\delta\Sigma(\bm{k})\ket{\bm{k},b},
    \label{eq:afm_sigma_response}
\end{align}
where $\ket{\bm{k},b}$ is the $b$th eigenvector of the self-consistent mean-field Hamiltonian at momentum $\bm{k}$, and $\delta\Sigma$ acts only on its conduction-electron block.
Thus a positive value of $-E_{\Sigma}^{\mathrm{AFM}}$ means that the occupied conduction-electron components of the antiferromagnetic quasiparticles overlap with momentum regions where the nonuniform self-energy lowers the energy.
Figure~\ref{fig:phase_filter_maps} shows the resulting phase responses.
In the top row, the longitudinal structure gives a positive $\Delta\mathcal{T}$, the transverse structure gives a negative $\Delta\mathcal{T}$, and the circular structure gives a much weaker response between them.
In the bottom row, the transverse structure gives a favorable antiferromagnetic self-energy contribution, whereas the longitudinal structure gives the opposite sign after the uniform part of $\Sigma(\bm{k})$ is removed.
The circular structure remains between these two limiting angular patterns.
An analogous projection onto the low-density FM phase, which is not shown, follows this trend, with the transverse structure most favorable, the longitudinal structure unfavorable, and the circular structure between them.

Figure~\ref{fig:phase_energy_difference} shows the relative phase energy $\delta E=E_{\mathrm{AFM}}-E_{\mathrm{Kondo}}$ at fixed $\kondo/t\simeq3.37$, obtained by interpolating the self-consistent energies to the transition point at zero cavity coupling.
For finite $\eta$, a positive value means that the heavy-fermion phase is lowered relative to the 120-degree antiferromagnetic phase, while a negative value means that the antiferromagnetic phase is lowered relative to the heavy-fermion phase.
The longitudinal type moves to the heavy-fermion side, the transverse type moves to the antiferromagnetic side, and the circular type produces a smaller shift between them.
Together, Figs.~\ref{fig:phase_filter_maps} and \ref{fig:phase_energy_difference} show that the numerical measures and the direct self-consistent energy comparison give the same polarization trend.

This trend can be understood from how each phase samples the self-energy map in Fig.~\ref{fig:sigma_bz}.
In the heavy-fermion phase, the relevant conduction-electron weight is distributed over the near-resonant momentum region that hybridizes with the nearly flat $f$ level.
The broad radial deformation produced by the longitudinal structure acts over this region and increases the low-energy conduction-electron weight available for Kondo hybridization.
This interpretation is consistent with the positive cavity-induced change $\Delta\mathcal{T}(\kondo ,\eta)$ in Fig.~\ref{fig:phase_filter_maps}.
The transverse structure instead concentrates the energy lowering in an annular region around the bare conduction-electron Fermi surface.
The numerical phase boundaries show that this structure gives a larger relative energy gain to both magnetic phases than to the heavy-fermion phase.
In the FM and AFM phases, the self-energy is sampled by spin-resolved and finite-momentum-mixed conduction-electron bands, respectively.
The detailed momentum distributions differ between the two magnetic phases, but the transverse structure is more favorable to both magnetic phases than the longitudinal structure.
The circular structure combines the two angular structures and lies between them.
This explains why the longitudinal type moves the transition toward smaller $\kondo$, in the same direction as the impurity trend where a cavity-induced mass enhancement strengthens the Kondo effect~\cite{mochidaCavityenhancedKondoEffect2024}.
Thus, the impurity mass-enhancement picture and the planar-cavity hyperbolic phonon-polariton dressing picture lead to the same expectation for the longitudinal projected-field type~\cite{ashidaCavityQuantumElectrodynamics2023}.
The connection between impurity mass enhancement and planar-cavity dressing is also consistent with the cavity and waveguide-QED intuition that strong light-matter dressing can renormalize electronic motion~\cite{ashidaNonperturbativeWaveguideQuantum2022,ashidaCavityQuantumElectrodynamics2021}.


\subsection{Estimate of Experimental Scales\label{sec:experimental_estimate}}
As a heuristic estimate, the main $\rho_{\mathrm e}=0.7$ calculation gives phase-boundary shifts up to about $4\%$ in the coupling range shown, depending on the polarization structure.
The half-filled triangular result in Appendix~\ref{app:filling} reaches about $9\%$.
A typical Kondo temperature is $T_{\mathrm{K}}\simeq3\,\mathrm{K}$~\cite{onukiAnisotropicMagneticProperty1985}, while dilution-refrigerator measurements on triangular TMD moir\'e materials have reached base temperatures near $20\,\mathrm{mK}$~\cite{ciorciaroKineticMagnetismTriangular2023}.
We therefore use $T_{\mathrm{K}}=3\,\mathrm{K}$ and $T=20\,\mathrm{mK}$ as illustrative scales below; this Kondo temperature is taken as a representative experimental scale rather than derived from the mean-field crossing at $\kondo^\ast(0)\simeq3.37t$.
We denote the critical Kondo coupling by $J_{\mathrm{K}}^{\mathrm c}(\eta)$ and define its cavity-induced shift as $\delta J_{\mathrm{K}}^{\mathrm c}(\eta)=J_{\mathrm{K}}^{\mathrm c}(\eta)-J_{\mathrm{K}}^{\mathrm c}(0)$.
Because the present mean-field transition is first order, the two competing energy branches cross with different slopes, and the displacement of the transition point produces a phase-selection energy that is linear in $|\delta J_{\mathrm{K}}^{\mathrm c}(\eta)|$.
We express the relative shift as $r(\eta)=|\delta J_{\mathrm{K}}^{\mathrm c}(\eta)|/J_{\mathrm{K}}^{\mathrm c}(0)$ and, taking the Kondo temperature as the characteristic low-energy scale, estimate this energy as $\Delta E_{\mathrm{cav}}\sim r(\eta)k_{\mathrm B}T_{\mathrm{K}}$.

For the h-BN-motivated longitudinal structure in the main calculation, the numerical results give $r(0.4)\simeq0.043$.
Since the leading cavity correction scales as $\eta^2$, we estimate $r(\eta)\sim0.043(\eta/0.4)^2$.
Requiring $\Delta E_{\mathrm{cav}}\gtrsim k_{\mathrm B}T$ then gives $\eta_{\min}\sim0.4\sqrt{T/(0.043T_{\mathrm{K}})}$.
For $T_{\mathrm{K}}=3\,\mathrm{K}$ and $T=20\,\mathrm{mK}$, where $k_{\mathrm B}T\simeq1.72\,\mu\mathrm{eV}$, this yields $\eta_{\min}\simeq0.16$, which we regard as a threshold of order $0.2$.
Therefore, we expect the cavity-induced modulation of the transition point to produce a detectable signal in experiments combining a two-dimensional TMD material with a hyperbolic polariton cavity.

\section{Conclusions\label{sec:conclusion}}
In this work, we studied a two-dimensional Kondo lattice confined in a planar dielectric cavity.
After identifying the cavity-induced electron--electron interaction, we included the leading static self-energy correction to the conduction-electron band and compared the heavy-fermion phase with the 120-degree AFM phase at the metallic filling $\rho_{\mathrm e}=0.7$ and with an assumed FM phase at low carrier density.
An h-BN hyperbolic phonon-polariton cavity provides a concrete motivation for the longitudinal projected field, while the transverse and circular in-plane structures can be considered as model extensions.

The leading cavity effect is a momentum-dependent dressing of the conduction-electron band.
The phase-boundary shifts reflect how each phase samples this self-energy through a different conduction-electron distribution.
The metallic AFM and low-density FM comparisons show a common trend.
The longitudinal h-BN-motivated field favors the heavy-fermion phase, while the transverse model structure favors the magnetic phases.
The circular structure lies between these two cases.
In particular, in the setting motivated by an h-BN cavity, the lattice model shows a coherent enhancement of the Kondo effect in the same direction as the impurity problem~\cite{mochidaCavityenhancedKondoEffect2024}.
The results can be tested in future cavity experiments where vacuum fluctuations are used to control  Kondo-lattice phase competition in gate-tunable moir\'e materials.
In real transition-metal dichalcogenide moir\'e systems, the conduction electrons and local moments may come from different layers, orbitals, or sublattices.
Our triangular Kondo model uses one effective conduction band and one spin-$1/2$ local moment in each moir\'e unit cell.
It provides a minimal description of the phase competition, and a device-specific description of MoTe$_2$/WSe$_2$ requires the corresponding layer, orbital, and sublattice structures.

The present result also points to a route for controlling quantum critical phenomena with the cavity vacuum fields.
In heavy-fermion and moir\'e materials, the low-energy phases are often determined by a small energy difference between itinerant and magnetic states.
Therefore, even a moderate cavity-induced change of the conduction-electron dispersion can be important when it acts in a momentum- and polarization-dependent way.
It should be interesting to extend this idea to other frustrated lattices, superconducting instabilities, and non-Fermi-liquid regimes near heavy-fermion quantum critical points.
Such studies will be useful for understanding how cavity QED can become a practical tool to control correlated quantum materials.

\begin{acknowledgments}
We are grateful to Kanta Masuki, Yiming Wang, Taiga Nakamoto and Takahiro Morimoto for fruitful discussions. J.M. acknowledges support from  Grant-in-Aid for JSPS Fellows (Grant No. JP26KJ0765). Y.A. acknowledges support from JST FOREST Program (Grant No. JPMJFR222U), JST CREST (Grant No. JPMJCR23I2), and JST [Moonshot R\&D] (Grant No. JPMJMS256J).
\end{acknowledgments}

\appendix

\section{Derivation of the Effective Interaction\label{app:effective_interaction}}

In this Appendix, we summarize how the cavity-induced electron--electron interaction used in the main text is obtained.
The derivation follows the perturbative projection onto the photon vacuum~\cite{masukiCavityMoireMaterials2024}.

The starting point is the minimal-coupling Hamiltonian for a two-dimensional conduction band in a periodic potential $V(\bm{r})$.
This should be regarded as an effective-mass continuum description before projecting to the lattice band used in the main text.
The localized $f$ orbitals are not written explicitly in this Appendix because we consider the situation where only the conduction electrons are coupled to the cavity field, while the local moments are not.
\begin{align}
    H &=
    \sum_{\sigma}\int d^2\bm{r}\,c_\sigma^\dagger(\bm{r})
    \left[
      \frac{(\bm{p}+e\bm{A}(\bm{r}))^2}{2m}+V(\bm{r})
    \right]c_\sigma(\bm{r})\notag\\
    &\quad +\sum_{\bm{q}}\hbar\omega_{\bm{q}} a^\dagger_{\bm{q}}a_{\bm{q}},
    \label{eq:app_minimal}
\end{align}
Expanding the kinetic term gives the matter Hamiltonian $H_0$, the paramagnetic interaction $H_I^{(A\cdot p)}$, and the diamagnetic $A^2$ term.
The $A^2$ term mainly renormalizes the photon frequency at this order and is not the source of the leading electron--electron interaction in the photon-vacuum projection used here.
Thus we focus on
\begin{align}
    H_I^{(A\cdot p)} &=
    \frac{e}{m}\sum_{\sigma}\int d^2\bm{r}\,
    c_\sigma^\dagger(\bm{r})
      \bm{A}(\bm{r})\cdot\bm{p}\,
    c_\sigma(\bm{r}).
    \label{eq:app_Ap}
\end{align}

We assume low-energy states of the product form
$|\psi\rangle=|\psi_0\rangle_{\mathrm{mat}}|0\rangle_{\mathrm{ph}}$.
The second-order perturbative correction through one virtual photon-polariton excitation gives an effective interaction between two conduction electrons.
For this step, we also assume that the electronic energy difference in the intermediate state is small compared with $\hbar\omega_{\bm{q}}$.

\begin{align}
    H_{\mathrm{eff}}^{(2)} &\simeq
    -\left(\frac{e\hbar}{m}\right)^2
    \sum_{\bm{k},\bm{k}',\bm{q},\sigma,\sigma'}|A_{\bm{q}}|^2
    \frac{(\bm{e}_{\bm{q}}\cdot\bm{k})(\bm{e}_{\bm{q}}^\ast\cdot\bm{k}')}{\hbar\omega_{\bm{q}}}
    \notag\\
    &\quad\times
    c^\dagger_{\bm{k}\sigma}c_{\bm{k}-\bm{q},\sigma}
    c^\dagger_{\bm{k}'+\bm{q},\sigma'}c_{\bm{k}'\sigma'}.
    \label{eq:app_second_order}
\end{align}

The factor $(e\hbar/m)^2$ comes from the paramagnetic $\bm{A}\cdot\bm{p}$ vertex in Eq.~\eqref{eq:app_Ap}.
In the low-energy approximation used here, the matrix element is $(e\hbar/m)(\bm{e}_{\bm{q}}\cdot\bm{k})A_{\bm{q}}$ for each photon vertex.
Normal ordering the fermionic operators produces the positive two-body term below, while the accompanying one-body contraction is absorbed into the reference dispersion.
The plane-wave normalization of the vector potential is taken as $A_{\bm{q}}^2\propto L^{-2}$, which cancels the area factor that appears when the momentum sum is written as a continuum integral.
Using $g_{\bm{q}}=eA_{\bm{q}}\sqrt{\omega_{\bm{q}}/(m\hbar)}$ and retaining the leading denominator $\hbar\omega_{\bm{q}}$, this interaction can be written in the form used in the main text,
\begin{align}
    H_{\mathrm{ee}} &=
    \sum_{\bm{q}}\frac{\hbar^2g_{\bm{q}}^2}{m\omega_{\bm{q}}^2}
    \sum_{\bm{k},\bm{k}',\sigma,\sigma'}
    (\bm{e}_{\bm{q}}\cdot\bm{k})(\bm{e}_{\bm{q}}^\ast\cdot\bm{k}')
    \notag\\
    &\quad\times
    c^\dagger_{\bm{k}+\bm{q},\sigma}c^\dagger_{\bm{k}'\sigma'}
    c_{\bm{k}'+\bm{q},\sigma'}c_{\bm{k}\sigma}.
    \label{eq:app_hee_final}
\end{align}
The momentum convention in Eq.~\eqref{eq:app_hee_final} is chosen to match Eq.~\eqref{eq:electron_electron} in the main text.
Different but equivalent conventions can be obtained by shifting $\bm{k}$ and $\bm{k}'$.
The direct cavity correction to the Kondo exchange is higher order in the hybridization expansion from the periodic Anderson model.
Since the usual Kondo exchange scales as $\kondo\sim V^2/U$, the cavity correction contains additional powers of the light-matter coupling and is neglected in the present treatment.

The effective coupling parameter $\eta$ and its Gaussian momentum profile are defined by
\begin{align}
   \frac{g_{\bm{q}}^2}{\omega_{\bm{q}}^2}
    = \frac{\eta^2}{D_0}\exp\left[
      -\frac{|\bm{q}-\bm{q}^\ast|^2}{2q_c^2}
    \right],
    \label{eq:app_vq}
\end{align}
where $D_0=\sum_{\bm q}\exp[-|\bm q-\bm q^\ast|^2/(2q_c^2)]$ uses the normalized momentum sum.
The self-energy contains this profile multiplied by the momentum-independent factor $\hbar^2/m$.
The numerical implementation absorbs this factor into the lattice energy unit and does not specify $g_{\bm{q}}$ and $\omega_{\bm{q}}$ separately.

Finally, the first-order Hartree--Fock self-energy from Eq.~\eqref{eq:app_hee_final} is
\begin{align}
    \Sigma_c^{(1)}(\bm{k}) =
    -2\sum_{\bm{q}}\frac{\hbar^2g_{\bm{q}}^2}{m\omega_{\bm{q}}^2}
    |\bm{e}_{\bm{q}}\cdot\bm{k}|^2f(\varepsilon_{\bm{k}-\bm{q}}),
    \label{eq:app_sigma_discrete}
\end{align}
where $f(\varepsilon)$ is the Fermi distribution function.
The overall factor 2 follows from the Fock contraction of the spin-summed interaction in Eq.~\eqref{eq:app_hee_final}, using the same convention as in Eq.~\eqref{eq:self_energy_intro}.
In the continuum notation this becomes Eq.~\eqref{eq:self_energy_intro}.
For the charge-uniform, zero-current mean-field states considered in the main text, the Hartree contraction is proportional to the conduction-electron current and vanishes after momentum summation.
It is therefore omitted in the numerical calculation.

\section{Additional Results for Other Lattices and Fillings\label{app:additional_results}}

In this Appendix, we present calculations that complement the analysis in the main text.
We examine the square lattice at half filling, the filling dependence on the triangular lattice, and the triangular lattice at half filling.

\subsection{Results for the Square Lattice\label{app:square}}
This subsection summarizes the corresponding mean-field results for a square-lattice Kondo model at half filling.
In this comparison, the magnetic phase is the N\'eel antiferromagnetic phase rather than the 120-degree antiferromagnetic phase.
The N\'eel order reconstructs the conduction band through two momentum sectors.
The phase labeled as the Kondo phase in this subsection was calculated as a coexisting mean-field state that allows a QSL-like bond field $Q$ in the $f$-electron sector together with the Kondo hybridization.

\begin{figure}[!htbp]
\centering
  \includegraphics[width=\linewidth]{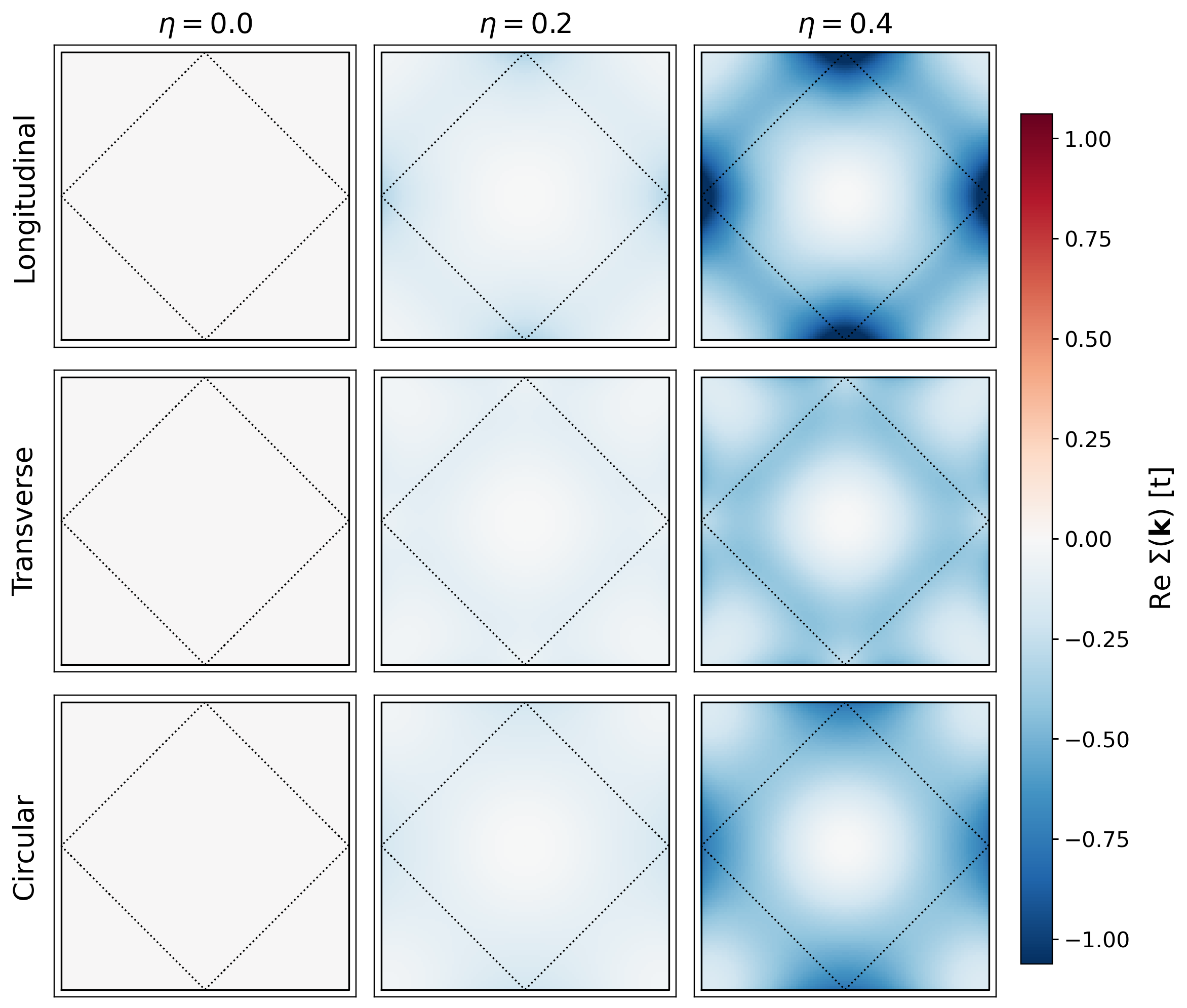}
  \caption{The self-energy $\Sigma(\bm{k})$ in momentum space for the square lattice at half filling.
  The columns show $\eta=0$, $0.2$, and $0.4$.
  The rows show the longitudinal, transverse, and circular polarizations.
  The black dotted line shows the bare conduction-electron Fermi surface.
  The longitudinal and transverse patterns are related by the square-lattice symmetry, and the circular case gives an averaged response.}
  \label{fig:square_sigma_bz}
\end{figure}

Figure~\ref{fig:square_sigma_bz} shows the square-lattice self-energy patterns.
The longitudinal and transverse patterns are nearly related by a square-lattice symmetry operation.

\begin{figure*}[!t]
\centering
  \includegraphics[width=0.88\textwidth]{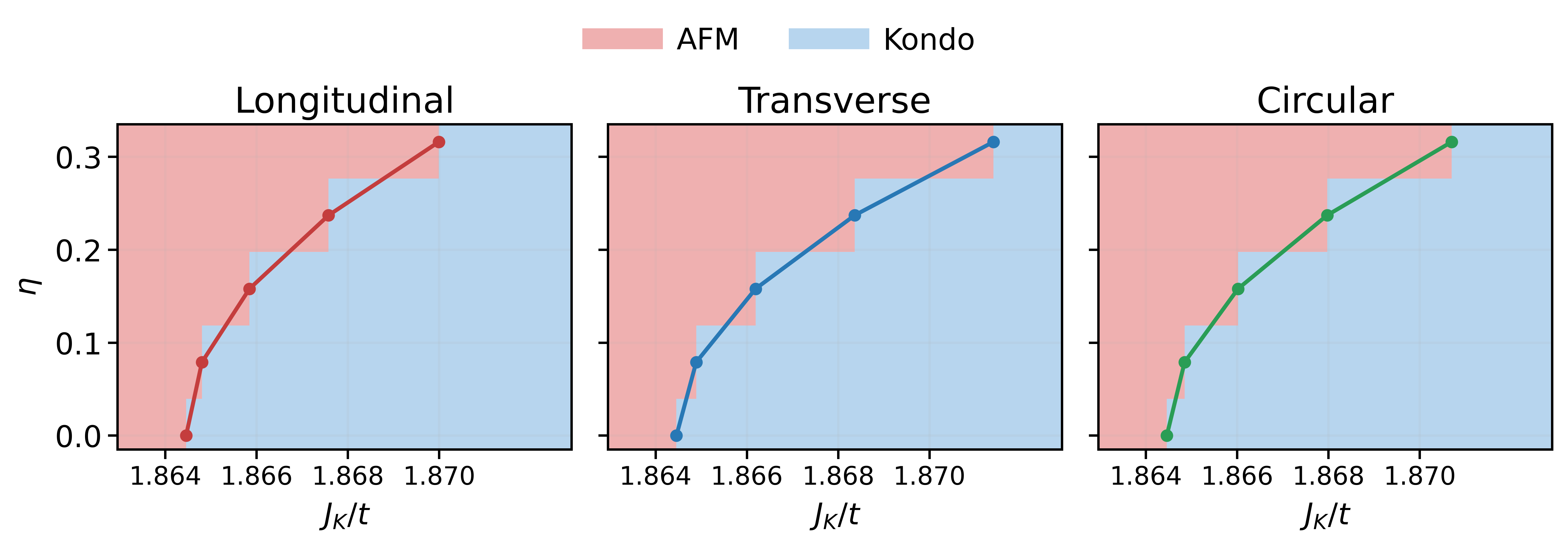}
  \caption{Square lattice phase diagram at half filling for the antiferromagnetic and heavy-fermion phases.
  The transition shifts only slightly to larger $\kondo$ as $\eta$ increases, and the polarization dependence is weaker than in the triangular lattice.}
  \label{fig:square_transition_eta}

\vspace{2mm}
\centering
  \includegraphics[width=0.42\textwidth]{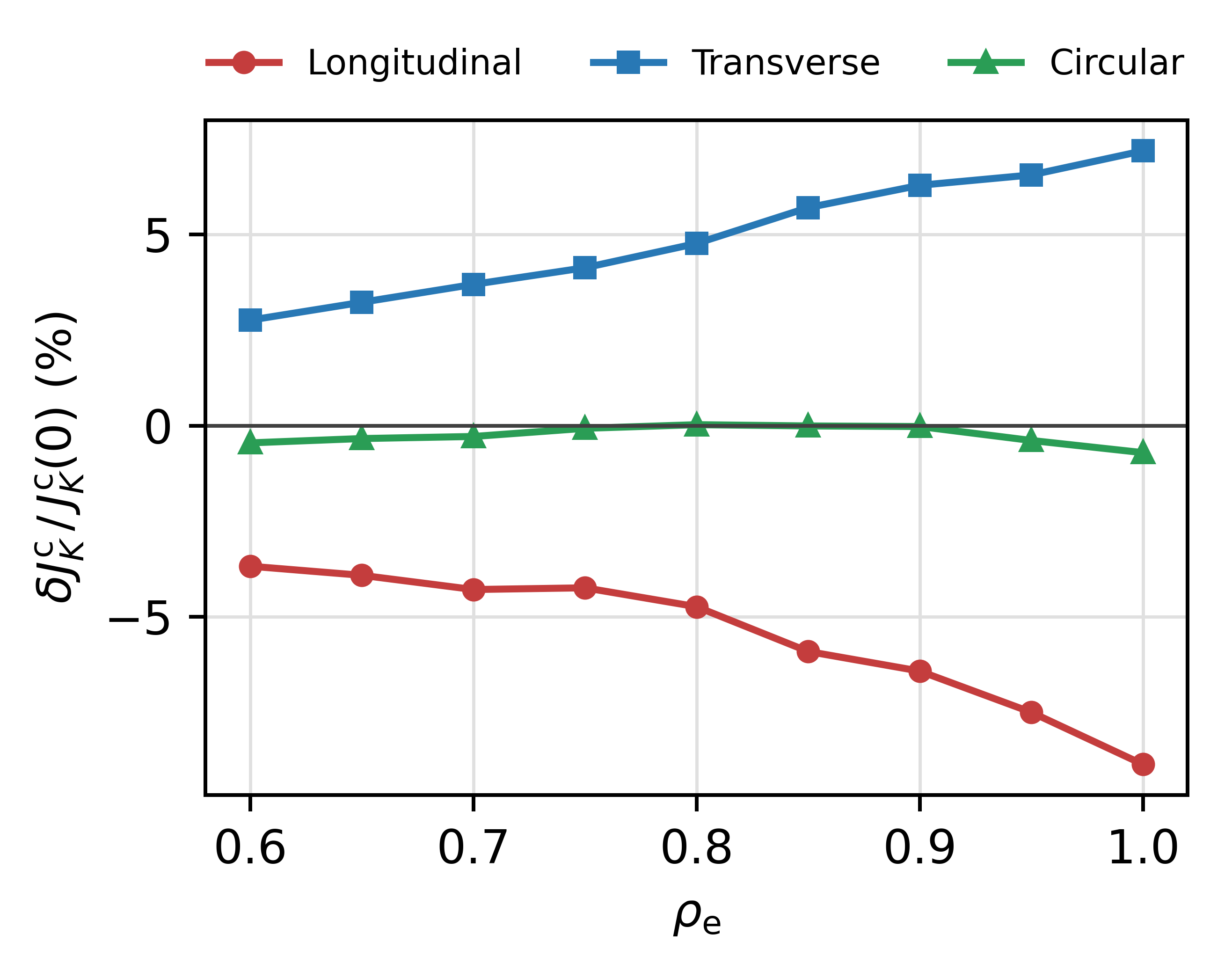}
  \caption{Relative shift of the triangular-lattice transition coupling at $\eta=0.4$ as a function of filling.
  The vertical axis shows $100\delta J_{\mathrm K}^{\mathrm c}(0.4)/J_{\mathrm K}^{\mathrm c}(0)$, where $\delta J_{\mathrm K}^{\mathrm c}(0.4)=J_{\mathrm K}^{\mathrm c}(0.4)-J_{\mathrm K}^{\mathrm c}(0)$.
  Negative values indicate that the heavy-fermion phase appears at a smaller Kondo coupling than at $\eta=0$.
  The magnitude of the cavity-induced shift generally increases as the filling approaches half filling.}
  \label{fig:filling_shift}

\vspace{2mm}
\centering
  \includegraphics[width=0.88\textwidth]{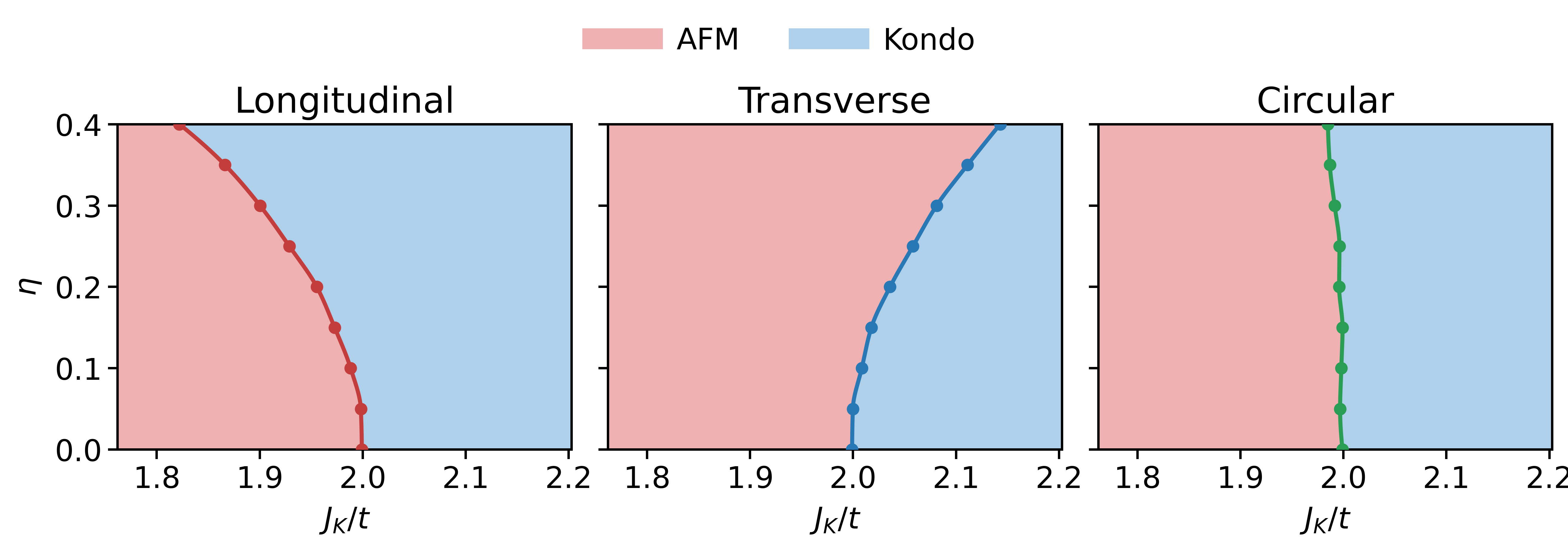}
  \caption{$J_{\mathrm{K}}$--$\eta$ phase diagram for the triangular lattice at half filling.
  The qualitative structure of the phase diagram is unchanged across the examined fillings.}
  \label{fig:tri_rho1_appendix}
\end{figure*}

Figure~\ref{fig:square_transition_eta} summarizes the corresponding phase diagram.
The transition shifts slightly to larger $\kondo$ as $\eta$ increases, and the three polarizations remain nearly degenerate.
This weak polarization dependence is consistent with the nearly symmetry-related self-energy patterns in Fig.~\ref{fig:square_sigma_bz}.
The two-sector N\'eel reconstruction samples these patterns and therefore produces only a modest difference among the polarization structures.

The heavy-fermion hybridization also has a weak polarization dependence in the square-lattice calculation.
Compared with the clear triangular-lattice polarization dependence discussed in Sec.~\ref{sec:discussion}, the change of the pure heavy-fermion hybridization $|P|$ is small for all polarizations.
The square-lattice phase boundary is therefore governed mainly by a weak, nearly common energy shift among the three polarization structures.

The corresponding energy comparison near the square lattice transition shows the same qualitative competition as in the triangular case, but the change of the crossing point remains modest even at large normalized $\eta$.

\subsection{Filling Dependence and the Half-Filled Triangular Lattice\label{app:filling}}

We analyzed how the triangular-lattice transition shift depends on filling and also calculated the half-filled phase diagram.
These results complement the metallic $\rho_{\mathrm e}=0.7$ phase diagram in the main text.
We also included the QSL-like spinon-bond mean field $Q$ that arises in large-$N$ theory, but all successful KondoQSL solutions in these scans converged to $Q=0$.
The density dependence discussed here should be regarded as a model calculation within a fixed effective reference band.
In an actual moir\'e material, the reference dispersion itself can change with carrier density because of ordinary Coulomb-induced band renormalization, and this density-dependent update of the reference band is not tracked explicitly in the present calculation.

Figure~\ref{fig:filling_shift} shows that the polarization dependence found at $\rho_{\mathrm e}=0.7$ persists over the examined filling range.
The longitudinal structure lowers the transition coupling, while the transverse structure raises it.
The circular structure gives a much smaller relative change.
The relative shifts generally become larger toward half filling for the longitudinal and transverse structures.
The larger relative shift toward half filling may arise because the expansion of the Fermi surface and occupied momentum distribution sample the momentum-dependent self-energy broadly.

Figure~\ref{fig:tri_rho1_appendix} shows the phase diagram at $\rho_{\mathrm e}=1$.
As at $\rho_{\mathrm e}=0.7$, the longitudinal structure lowers the transition coupling, the transverse structure raises it, and the circular structure produces a much smaller change.
This trend holds for all examined fillings with $\rho_{\mathrm e}>0.5$.

\clearpage

\bibliography{lattice}

\end{document}